\documentclass[10pt,journal,compsoc,onecolumn]{IEEEtran}
\ifCLASSOPTIONcompsoc
  \usepackage[nocompress]{cite}
\else
  \usepackage{cite}
\fi
\usepackage{subfigure,cite,graphicx,amsmath,algorithmic,amssymb,mathrsfs,epsf}
\usepackage[usenames]{color}
\usepackage{multirow}
\usepackage{tabularx}
\usepackage{fixltx2e}
\usepackage{epstopdf}
\usepackage{textcomp}
\usepackage{tikz}
\usepackage{ctable}
\usepackage{lipsum}
\usepackage{enumitem}
\usepackage{amsthm}
\usepackage{verbatim}
\usepackage{xcolor}
\usetikzlibrary{calc}
\usepackage{soul}
\usepackage{pgfplots}
\usepackage{stfloats}
\usepackage{setspace}
\usepackage{caption}
\usepackage{algorithm}
\newlength\myindent
\newcommand\bindent{%
  \begingroup
  \setlength{\itemindent}{\myindent}
  \addtolength{\algorithmicindent}{\myindent}
}
\newcommand\eindent{\endgroup}
\begin{document}
\sloppy
\title{How to Cache in Mobile Hybrid IoT Networks?}
\author{Trung-Anh Do, Sang-Woon Jeon,~\IEEEmembership{Member,~IEEE,} \\ and
        Won-Yong Shin,~\IEEEmembership{Senior Member,~IEEE}
\thanks{The
material in this paper was presented in part at the IEEE
International Symposium on Information Theory, Barcelona, Spain,
July 2016~\cite{c7}. (\textit{Corresponding author: Won-Yong
Shin}.)}
\thanks{T.-A. Do was with
Dankook University, Yongin 16890, Republic of Korea. He is now
with the Division of Science and Technology Management and
International Cooperation, Posts and Telecommunications Institute
of Technology, Hanoi 100000, Vietnam (e-mail: anhdt@ptit.edu.vn).}
\thanks{S.-W. Jeon is with the Department of
Military Information Engineering, Hanyang University, Ansan 15588,
Republic of Korea (e-mail: sangwoonjeon@hanyang.ac.kr).}
\thanks{W.-Y. Shin is with the Department of
Computer Science and Engineering, Dankook University,
Yongin 16890, Republic of Korea (e-mail: wyshin@dankook.ac.kr).}}

\newtheorem{definition}{Definition}
\newtheorem{thm}{Theorem}
\newtheorem{lemma}{Lemma}
\newtheorem{example}{Example}
\newtheorem{corollary}{Corollary}
\newtheorem{proposition}{Proposition}
\newtheorem{conjecture}{Conjecture}
\newtheorem{remark}{Remark}

\newcommand{\red}[1]{{\textcolor[rgb]{1,0,0}{#1}}}

\def \diag{\operatornamewithlimits{diag}}
\def \min{\operatornamewithlimits{min}}
\def \max{\operatornamewithlimits{max}}
\def \log{\operatorname{log}}
\def \max{\operatorname{max}}
\def \rank{\operatorname{rank}}
\def \out{\operatorname{out}}
\def \exp{\operatorname{exp}}
\def \arg{\operatorname{arg}}
\def \E{\operatorname{E}}
\def \tr{\operatorname{tr}}
\def \SNR{\operatorname{SNR}}
\def \dB{\operatorname{dB}}
\def \ln{\operatorname{ln}}
\def \bmat{ \begin{bmatrix} }
\def \emat{ \end{bmatrix} }
\def \be {\begin{eqnarray}}
\def \ee {\end{eqnarray}}
\def \ben {\begin{eqnarray*}}
\def \een {\end{eqnarray*}}

\newcommand{\Pro}[1]{\mathrm{Pr}\left\{#1\right\}}
\newcommand{\LIF}[2]{\tilde{L}_{\pi_1(#1),#2}}
\newcommand{\TIL}[2]{L_{\pi_2(#1),#2}}
\newcommand{\TIF}[2]{T_{\pi_1(#1),#2}}
\newcommand{\KIF}[2]{T_{\pi_1(#1),\pi_2(#2)}}
\newcommand{\snr}{\textsf{snr}}
\newcommand{\sinr}{\textsf{sinr}}
\newcommand{\CanSB}{\mathcal{B}}
\newcommand{\CanSA}{\mathcal{A}}
\newcommand{\Norm}[1]{\left|{#1}\right|}

\IEEEtitleabstractindextext{
\begin{abstract}
Content-centric {\em mobile hybrid} Internet-of-Things (IoT) networks
consisting of mobile devices and static femto access points (FAPs)
are studied, where each device moves
according to the random walk mobility model and requests a content
object from the library independently at random according to a
Zipf popularity distribution. Instead of allowing access to
content objects at macro base stations via costly backhaul providing
connection to the core network, we consider a more practical scenario where mobile devices and static FAPs, each having a
{\em finite-size} cache space, are able to cache a subset of
content objects so that each request is served by other mobile
devices or static FAPs. Under a general multihop-based content delivery protocol, we analyze the order-optimal
throughput--delay trade-off by presenting a new cache allocation
strategy. In particular, under a
given caching strategy, we first characterize a throughput--delay trade-off in terms of scaling laws along with the general content delivery multihop routing protocol.
Then, the order-optimal throughput--delay trade-off is characterized by
presenting the order-optimal cache allocation strategy, which jointly
finds the replication sets at mobile devices and static FAPs via a novel
variable decoupling approach. In our mobile IoT network, an interesting observation is that
highly popular content objects are mainly served by mobile devices
while the rest of content objects are served by static FAPs. We
perform numerical evaluation to validate our analytical results.
We also show that the order-optimal strategy strictly outperforms a
baseline approach, where the replication sets at mobile devices and static FAPs are optimized separately.
\end{abstract}
}
\maketitle
\IEEEdisplaynotcompsoctitleabstractindextext

%
\IEEEpeerreviewmaketitle
\section{Introduction}
Wireless data
caching~\cite{singlehop} has emerged as a promising technique
that effectively deals with the exponential growth of data traffic caused by mobile Internet-of-Things (IoT) devices~\cite{surv,acc1,acc2} without introducing costly backhaul (or infrastructure) providing
connection to the core network, while maintaining the sustainability of future wireless networks. 
The core of wireless
data caching in content-centric IoT networks is to allow base stations or end terminals to cache a subset of content objects. Hence, user requests can be
directly served by base stations or end terminals that have cached the requested objects, without contacting costly backhaul links.

\subsection{Prior Work}
As the number of users continues to grow dramatically, the
capacity scaling law behavior has been widely studied in
large-scale IoT networks. Gupta and Kumar showed in \cite{c8} that for
a static IoT network consisting of $n$ randomly distributed
source--destination (S--D) pairs in a unit network area, the
per-device throughput of $\Theta\left(\frac{1}{\sqrt{n \log
n}}\right)$ is achievable using the nearest neighbor multihop
transmission. There have been further studies on multihop schemes
in the literature~\cite{Franceschetti,GuptaKumar2003,Xue,Shin},
where the per-device throughput scales far slower than $\Theta(1)$.
Besides the multihop schemes, there have been various research
directions to improve the per-device throughput up to a constant
scaling by using hierarchical cooperation~\cite{c11}, device mobility~\cite{c12,c1}, directional
antennas~\cite{Zhang,Li,Yoon}, and infrastructure
support~\cite{Liu,Shin11}.

Contrary to the studies on the conventional IoT network model
in which S--D pairs are given and fixed, investigating
\textit{content-centric IoT networks} would be quite
challenging. As content objects are cached by numerous mobile IoT devices over
a network, finding the closest content holder of each request and
scheduling between requests are of crucially importance for
improving the overall network performance. The scaling behavior of
content-centric IoT networks has received a lot of attention in
the literature~\cite{c2,Ji_IT,c5,c3,a1,XLiu}. In \textit{static} ad
hoc networks, throughput scaling laws were analyzed using multihop
communication~\cite{c2,c5}, which yields a significant performance
gain over the single-hop caching scenario
in~\cite{singlehop,Ji_IT}. More specifically, a centralized and
deterministic cache allocation strategy was presented
in~\cite{c2}, where replicas of each content object are statically
determined based on the popularity of each content object in a
centralized manner. A decentralized and random cache allocation
strategy along with a local multihop protocol was also introduced
in~\cite{c5}, where content objects are assigned independently at
random to the caches of all users. On the other hand, in
\textit{mobile} IoT networks, performance on the throughput and
delay was examined under a reshuffling mobility model, where the
position of each mobile device is independently determined according to
random walks with an adjustable flight size and updated at the
beginning of each time slot~\cite{c3}. Alfano et al. showed in ~\cite{c3}
that increasing the mobility degrees of mobile devices leads to worse
performance when deterministic cache allocation is used similarly
as in~\cite{c2}. In~\cite{Adeel}, the above performance analysis was then extended to the case where the size of each content object is considerably large and thus
only a subpacket of a file can be delivered during one time slot.
Performance on the throughput and delay was also investigated in \cite{XLiu} under a correlated mobility model, where mobile devices are partitioned into multiple clusters and the devices belonging to the same cluster move in a correlated fashion. Liu et al. showed in \cite{XLiu} how correlated mobility affects the network performance. In addition, caching in IoT networks was extended to {\em static} infrastructure-supported IoT networks using multihop communication~\cite{acc3,a1}---each macro base station was assumed to be connected to the core network via infinite-speed backhaul, which has an access to all content objects stored in the
whole file library.

Meanwhile, a different caching framework, termed coded
caching~\cite{Nie,Nie1,Nie3}, has received a great deal of
attention in cache-enabled wireless networks, where a single
transmitter simultaneously deals with several different demands
using common coded multicast transmission. Based on this approach,
a global caching gain can be achieved by finding the optimal
content placement such that multicasting opportunities are
exploited simultaneously for all possible requests in the delivery
phase.

\subsection{Main Contributions}
In this paper, we study a large-scale content-centric {\em mobile hybrid multihop IoT} network, where each mobile device moves according to the random walk mobility model (RWMM) and requests a content object from
the library independently at random according to a Zipf popularity
distribution while multiple femto access points (FAPs) (or helper devices) are
regularly placed over the network area. 
Instead of assuming an access to the core network through
infinite-speed backhaul links as in \cite{a1},
we assume that \textit{each of mobile devices and FAPs is equipped
with a finite-size cache} and is able to cache content objects in
the library. Our caching
framework is basically composed of the caching phase and the
delivery phase. For a given caching strategy, we first present a
content delivery routing protocol with and without FAP support via
multihop in order to deliver content objects to requesting mobile devices,
which leads to a fundamental throughput--delay trade-off. Then, we
optimize a cache allocation strategy that provides the
order-optimal throughput--delay trade-off. Specifically, we propose a novel \textit{variable
decoupling} approach that optimally finds the replication sets for
caching at mobile devices and static FAPs, denoted by $A_m$ and $B_m$ for
content object $m\in\{1,\cdots,M\}$, respectively, where $M$
indicates the number of content objects in the library. 
The proposed approach solves two different convex optimization problems with relaxation based on the relative size of $A_m$ and $B_m$, leading to much simpler analysis without loss of order optimality compared to tackling the original problem that may not provide a tractable closed-form
solution. In our mobile IoT network, main results reveal that when each FAP has a
relatively large-size cache, highly popular content objects are
mainly served by mobile devices whereas the rest of content objects
are served by static FAPs. Based on the order-optimal cache
allocation strategy, we finally characterize the order-optimal
throughput--delay trade-off with respect to system parameters.

The main contributions of this paper are summarized
as follows:
\begin{itemize}
\item For comprehensive understanding of content-centric IoT networks, we present a general framework in which both static FAPs and mobile devices are able to cache a subset of content objects with different capabilities, also capturing the effects of user mobility and multihop content delivery. 

\item Under such a general setting, the order-optimal  throughput--delay trade-off is analyzed, which is the first result that characterizes a fundamental trade-off between throughput and delay of content-centric mobile IoT networks.

\item The main technical challenge resides in establishing the order-optimal content
replication strategy (i.e., the order-optimal cache allocation
strategy), which jointly optimizes the number of replicas cached
at mobile devices and static FAPs, denoted by $A_m$ and $B_m$ for
$m\in\{1,\cdots,M\}$, respectively. An interesting observation is that when the total cache
space at all FAPs is greater than that at all mobile devices, highly
popular contents are stored mainly in mobile device caches while any
request for less popular contents is fulfilled by static FAPs.


\item Our analytical solution for the order-optimal caching placement is compared with a numerically optimized solution, demonstrating that our  scaling law analysis is well matched with the numerical results.

\item For comparison, a baseline strategy that
optimizes the replication sets at mobile devices and static FAPs in a
separate manner is further presented and shows that it is strictly suboptimal.
\end{itemize}

\subsection{Organization}
The rest of this paper is organized as follows. In Section II, the
network model and performance metrics under consideration are
described. In Section III, the content delivery routing protocol
is presented. In Section IV, a fundamental throughput--delay
trade-off is introduced in terms of scaling laws. In Section V,
the order-optimal throughput--delay trade-off is derived by introducing
the order-optimal cache allocation strategy using variable decoupling. In Section VI, numerical results are shown for validation. A
baseline strategy is also shown in Section VII for comparison.
Finally, Section VIII summarizes the paper with some concluding
remarks.

\subsection{Notations}
Throughout this paper, $\mathbb{E}\left[\cdot\right]$  and
$\Pr\left(\cdot\right)$ are the expectation and the probability,
respectively. Unless otherwise stated, all logarithms are assumed
to be to the base 2. We also use the following asymptotic
notation: i) $f(x)=O(g(x))$ means that there exist constants $C$
and $c$ such that $f(x)\leq Cg(x)$ for all $x>c$; ii)
$f(x)=o(g(x))$ means that $\lim_{x\rightarrow \infty}
\frac{f(x)}{g(x)}=0$; iii) $f(x)=\Omega(g(x))$ if $g(x)=O(f(x))$;
iv) $f(x)=\omega(g(x))$ if $g(x)=o(f(x))$; and v)
$f(x)=\Theta(g(x))$ if $f(x)=O(g(x))$ and
$f(x)=\Omega(g(x))$~\cite{c10}. That is, $f(x)=O(g(x))$ means that $g(x)$ increases faster than or equal to $f(x)$ in an order sense. Similarly, $f(x)=o(g(x))$  means that $g(x)$ increases strictly faster than $f(x)$ in an order sense. Lastly, $f(x)=\Omega(g(x))$ means that $f(x)$ and $g(x)$ increase with the same order.

\section{Network Model and Performance Metrics} \label{SEC:model}
In this section, we first describe the network model and then define performance metrics used in the paper.

\subsection{Network Model}
We consider a content-centric mobile hybrid IoT network consisting of
$n$ mobile devices and $f(n)=\Theta(n^{\delta})$ static FAPs (or
static helper devices), where $0\leq\delta<1$, which is a general network model including the prior models \cite{singlehop,c1,c2,Ji_IT,c5,c3,a1}.
We assume that $n$ mobile devices are distributed uniformly at random over a unit square and $f(n)$ FAPs are regularly placed over the same area. That is, the network is divided into $f(n)$ square femto cells of equal size given by $b(n)=\Theta\left(\frac{1}{f(n)}\right)$ so that each cell has one FAP at its center. 
The mobility trace of devices is modelled according to the RWMM as in~\cite{c1,RW2}. In particular, the unit area is divided into $n$ square subcells of area $\frac{1}{n}$. 
Each mobile device independently performs a simple random walk with a
distance $\frac{1}{\sqrt{n}}$ on the $\sqrt{n}\times\sqrt{n}$
disjoint subcells so that the mobile device is equally likely to be in any
of the four adjacent subcells after each time slot.

In our content-centric mobile IoT network, each mobile device and static FAP are assumed to be equipped with local caches, which are installed
to store a subset of content objects in the central server with a database of
$M=\Theta(n^{\gamma})$ content objects, where $0 < \gamma<1$. Every content object
is assumed to have the same size. In this paper, we consider a
more practical cache-enabled network model by assuming that each
device and FAP are able to cache at most $K_n=\Theta(1)$ and
$K_{FAP}=\Theta(n^{\beta})$ content objects in their own
finite-size caches, respectively, where $0 < \beta<\gamma$. We
focus on the case that the total cache size in static FAPs scales no
slower than the total cache size of mobile devices in the network,
i.e., $\delta+\beta \geq 1$, in order to analyze the impact and
benefits of FAPs equipped with a relatively large-size
cache.\footnote{Otherwise, the use of FAPs would not be beneficial
in improving performance on the throughput and delay in our
network model.}

We assume that every mobile device requests its content object
independently according to a Zipf popularity distribution, which
typically characterizes a popularity of various kinds of real data
such as web, file sharing, user-generated content, and video on
demand \cite{c6}.\footnote{Note that a
Zipf popularity has also widely been adopted in mobile
IoT networks~\cite{Zipf,Zipf1}.} That is, the request probability of
content object $m \in \mathcal{M}\triangleq \left\lbrace 1,\cdots,M
\right\rbrace $ is given by\footnote{Without loss of generality,
we assume a descending order between the request probabilities of
the $M$ content objects in the library.}
\begin{equation}
\label{eq1}
p_m=\frac{m^{-\alpha}}{H_{\alpha}(M)},
\end{equation}
where $\alpha>0$ is the Zipf exponent and $H_{\alpha}(M)=\sum_{i=1}^M i^{-\alpha}$ is a normalization constant formed in the Riemann zeta function and is given by
\begin{equation}
\label{equa9} H_{\alpha}(M) =
  \begin{cases}
   \Theta(1)& \mbox{for } \alpha>1 \\
   \Theta(\log M) & \mbox{for }\alpha=1 \\
      \Theta(M^{1-\alpha}) & \mbox{for }\alpha<1.
  \end{cases}
\end{equation}


In content-centric networks, a caching problem is generally
partitioned into the caching phase and the delivery phase. That
is, the problem consists of storing content objects in the caches
and establishing efficient delivery routing paths for the
requested content objects.

We first consider the caching phase, which takes place during off-peak periods to proactively selects the content objects from the central server to be stored in the caches of $n$ devices and $f(n)$ FAPs. Let $A_m$ and $B_m$ denote the number of replicas of content
object $m \in \mathcal{M}$ stored at mobile devices and static FAPs, respectively,
which will be optimized later. In order for a cache allocation to
be feasible, $\{A_m\}_{m=1}^M$ and $\{B_m\}_{m=1}^M$ should
satisfy the following total caching constraints:
\begin{equation}
\label{equa1}
\begin{cases}
   \sum_{m=1}^{M} A_m \leq nK_n, \\
   \sum_{m=1}^{M}B_m \leq f(n)K_{FAP}.
  \end{cases}
\end{equation}
Furthermore, we impose the following individual caching constraints:
\begin{equation}
\label{equa01}
\begin{cases}
    A_m \leq n,  \\
    B_m \leq f(n),  \\
  A_m+B_m \geq 1
  \end{cases}
\end{equation}
for all $m \in \mathcal{M}$. Note that the last constraint in
\eqref{equa01} is needed to avoid an outage event such that a
requested content object is not stored in the entire network.
Similarly as in \cite{c2,c3}, we employ the \textit{random
caching} strategy such that the sets of replicas satsfying
\eqref{equa1} and \eqref{equa01} are stored uniformly at random
over the caches in $n$ devices and $f(n)$ FAPs.

Now, let us consider the delivery phase of the requested content
objects, which allows the requested content objects to be
delivered to the corresponding mobile devices over wireless channels.
During the delivery phase, each mobile device downloads its requested
content object (possibly via multihop) from one of the mobile devices or
static FAPs storing the requested content object in their caches. We
adopt the \textit{protocol model} \cite{c8} for successful content
delivery. In particular, let $d(u,v)$ denote the Euclidean
distance between devices $u$ and $v$. Then, content delivery from
device $u$ to device $v$ is assumed to be successful if and only if
$d(u,v) \leq r$ and there is no other active transmitter in a
circle of radius $(1+\Delta)r$ from device $v$, where $r$ and
$\Delta>0$ are given protocol parameters. \par For analytical
tractability, we also adopt the {\em fluid model} in \cite{c1}. In this
model, the size of each content object is assumed to be
arbitrarily small. Accordingly, the time required
for delivery of one content between a device and its neighbor device
or an assigned FAP is much smaller than the duration of each time
slot. In this case, the data sent from a device in
one time slot may correspond to multiple content objects, and
thus all content objects waiting for transmission at a device will
be transmitted by the device within one time slot. However, a
content object received by a device in a given time slot cannot be
transmitted by the device until the next time slot.

\subsection{Performance Metrics}
Since every content object is assumed to have the same size, we
consider the per-device throughput $\lambda(n)$, i.e., the rate at
which the request of any mobile device in the network can be served
according to a given feasible content delivery routing. We make
slight modifications to the definitions of throughput and delay in
\cite{c1} to fit into our content-centric mobile IoT network,
which are provided as follows.
\begin{definition}[Throughput] Let $B(i,t)$ denote the total number of bits of the requested content objects received by device $i$ during $t$ time slots. Note that this could be a random quantity for a given network realization.
Then, the per-device throughput $\lambda(n)$ is said to be achievable if there exists a sequence of events $A(n)$ such that
\begin{equation}
A(n)  =\left\lbrace \min_{1\leq i \leq n} \liminf\limits_{t\rightarrow \infty} \frac{1}{t} B(i,t) \geq \lambda(n)\right\rbrace \nonumber
\end{equation}
and $\Pr\left(A(n)\right)$ approaches one as $n$ tends to infinity.
\end{definition}
\begin{definition}[Delay] Let $D(i,k)$ denote the delay of the $k$th requested content object of device $i$, which is measured from the moment that the requesting message leaves device $i$ until the corresponding content object arrives at the device from the closest holder. For a particular realization of the network, the delay for device $i$ is $\limsup\limits_{q\rightarrow \infty} \frac{1}{q} \sum_{k=1}^q D(i,k)$ for a sufficiently large number of requested content objects of the device.
Then, the delay is defined as the expectation of the average delay of all devices over all network realizations, i.e.,
\begin{equation}
D(n) \triangleq \mathbb{E}\left[\frac{1}{n} \sum_{i=1}^n \limsup\limits_{q\rightarrow \infty} \frac{1}{q} \sum_{k=1}^q D(i,k)\right]. \nonumber
\end{equation}
\end{definition}

\section{Content Delivery Routing Protocol in Mobile Hybrid IoT Networks} \label{routing}
In this section, we describe our routing protocol to deliver
content objects to requesting mobile devices. Due to the device mobility, our
routing protocol is basically built upon the nearest neighbor
multihop routing scheme in~\cite{c1} and reconstructed for our
cache-enabled setting accordingly. For multihop transmission, the
network of unit area is divided into $a(n)^{-1}$ square routing
cells of equal size, where $a(n)=\Omega\left(\frac{\log
n}{n}\right)$ and $a(n)=O(1)$, so that each routing cell has at
least one mobile device with high probability (whp) (see \cite{c8} for the details). 
\begin{figure*}[t!]
    \centering
     \subfigure[Device to device]{
  \includegraphics[width=0.48\textwidth]{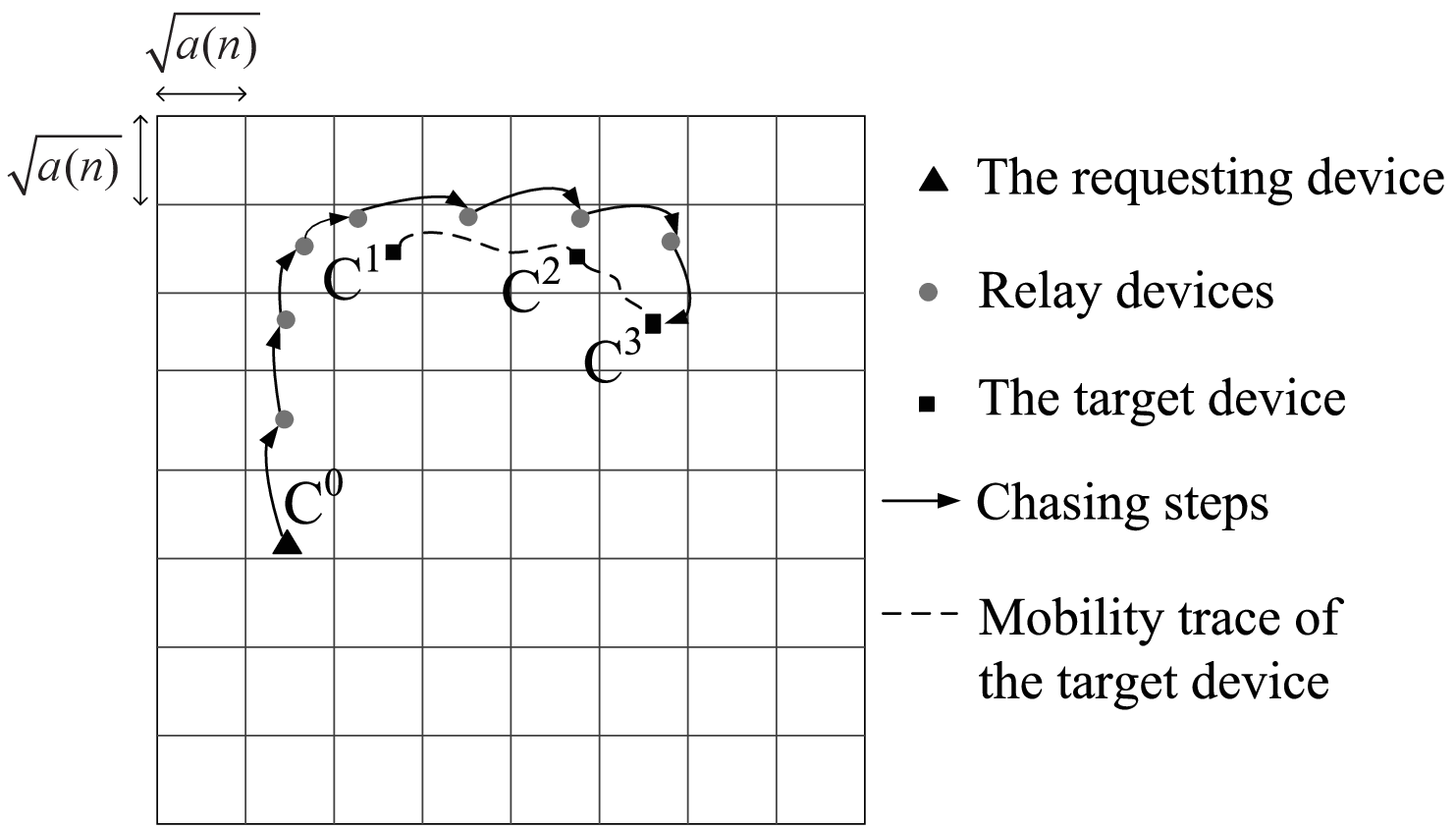}
       \label{1.1}
}
       \subfigure[Device to FAP]{
  \includegraphics[width=0.48\textwidth]{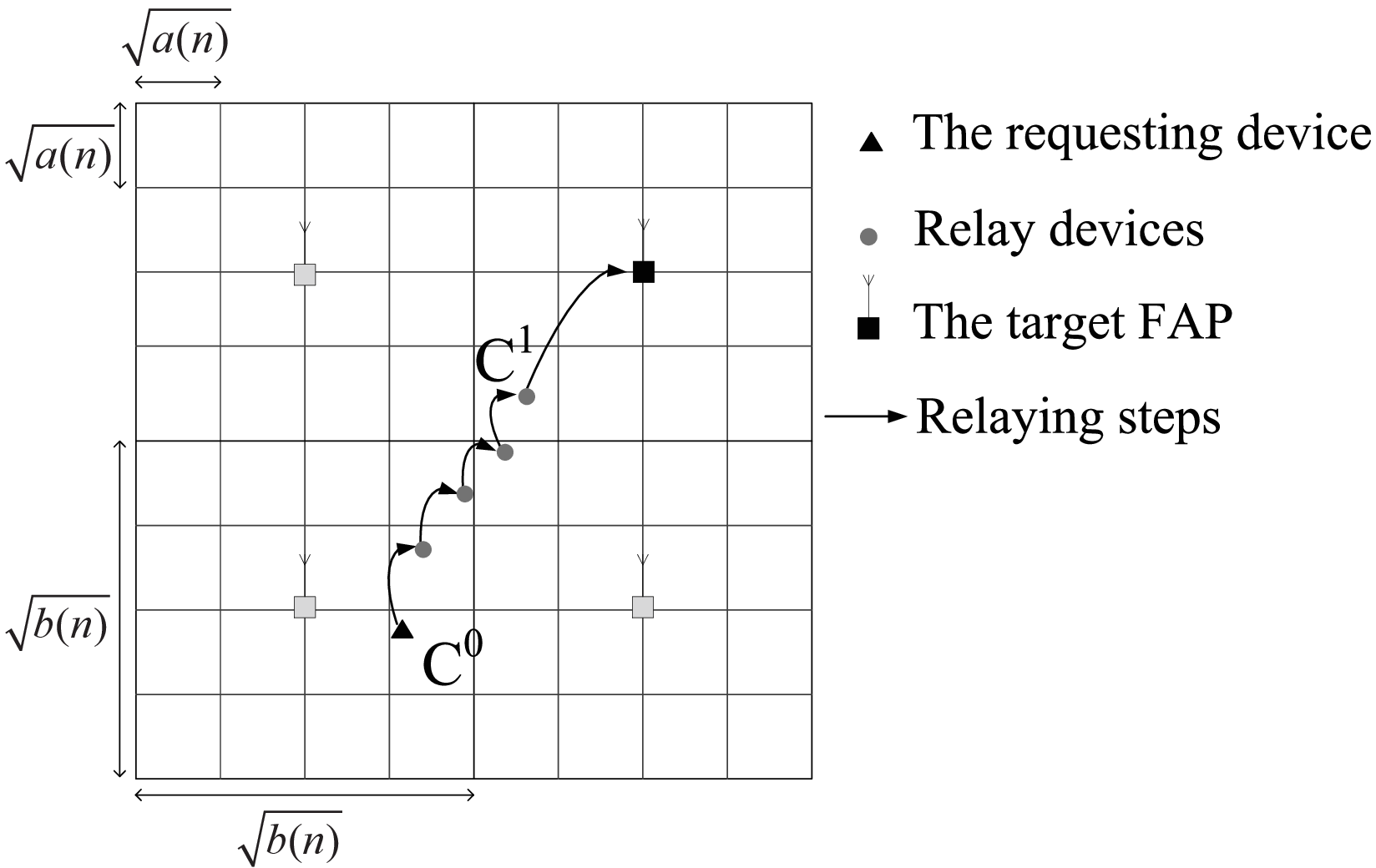}
        \label{1.2}
 }
    \caption{\small{The first phase of the content delivery routing.}}
    \label{1}
\end{figure*}
We implement a multihop routing strategy for content delivery
based on {\em routing cells} and {\em femto cells} whose size is $a(n)$ and
$b(n)$, respectively. Each routing cell is activated regularly
once every $1+c$ time slots to avoid any collision, where $c>0$
denotes a small integer independent of $n$. Similarly, each femto
cell is activated regularly once every $1+c$ time slots. 

The requesting mobile device first finds its closest holder (in the
Euclidean distance) of the desired content object among $A_m$
devices and $B_m$ FAPs. Then, a requesting message is delivered to
the closest holder along the adjacent routing cells via multihop
in forward direction, which corresponds to the first phase of the
content delivery. Similarly, the desired content object chases the
requesting device moving according to the RWMM via multihop in
backward direction, which corresponds to the second phase. Each
time slot is divided into two sub-slots. The first and second
phases of the content delivery procedure are activated during the
first and the second sub-slots, respectively. For the case where
the requesting mobile device is inside the transmission range of any holder
of the desired content object, the request will be served using
single-hop transmission within one time slot.
The detailed content delivery
procedure is described as follows:
\begin{figure*}[t!]
    \centering
     \subfigure[Device to device]{
  \includegraphics[width=0.5\textwidth]{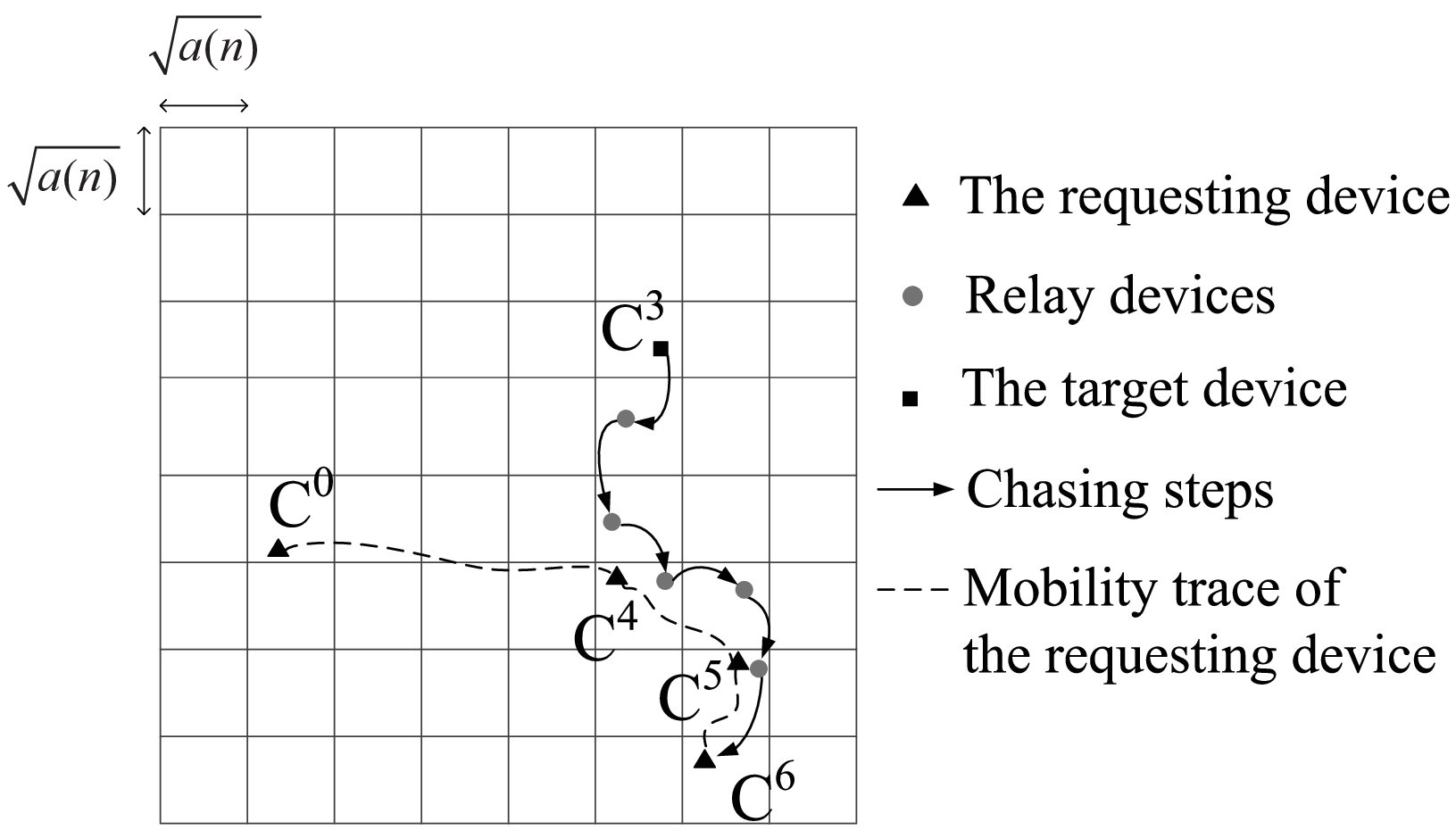}
       \label{2.1}
}
       \subfigure[FAP to device]{
  \includegraphics[width=0.48\textwidth]{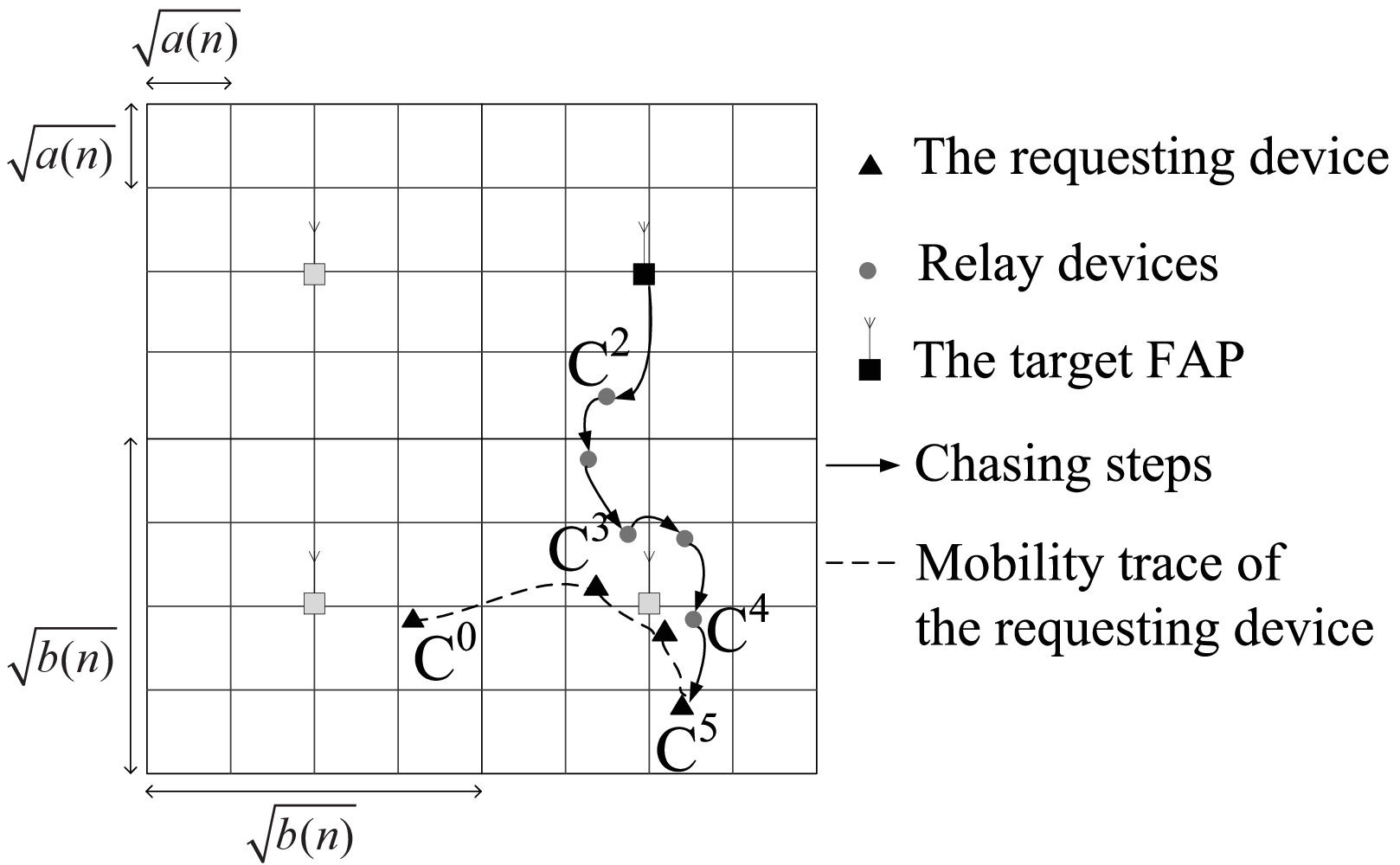}
        \label{2.2}
 }
    \caption{\small{The second phase of the content delivery routing.}}
    \label{2}
\end{figure*}

\begin{enumerate}[leftmargin=0.325cm, labelsep=0.325cm, align=left, itemsep=-0.1cm, font=\normalfont, label=(\roman*)]
\item[\textbf{Step 1)}] \underline{Requesting phase}
\begin{enumerate}
\item If the closest holder is a mobile device, then the requesting
message chases the target device according to the following
procedure. As depicted in Fig. \ref{1.1}, from routing cell
$C^{0}$, the requesting message is transmitted via multihop along
the adjacent routing cells toward routing cell $C^{1}$ containing
the target device, where the per-hop distance is given by
$\Theta\left(\sqrt{a(n)}\right)$. By the time the requesting
message reaches routing cell $C^{1}$, the target device has moved to
another position $C^{2}$ with the mobility trace based on the
RWMM. Thus, the message hops from routing cell $C^{1}$ to routing
cell $C^{2}$. This continues until the message reaches the routing
cell  $C^{3}$ containing its target device. \vspace{0.15cm} \item As
depicted in Fig. \ref{1.2}, if the closest holder is an FAP, then
the requesting message is delivered via multihop along the
adjacent routing cells intersecting the straight line toward the
coverage of the target FAP, where the per-hop distance is given by
$\Theta\left(\sqrt{a(n)}\right)$. When the requesting message
arrives at a mobile device in routing cell $C^{1}$ that is inside the FAP
cell coverage of the target FAP, the last relay device will send the
message to the FAP immediately using single-hop within one time
slot, where the last hop distance to the target FAP is given by
$\Theta\left(\sqrt{b(n)}\right)$. This long-distance hop for the
last hop to the FAP leads to the best performance (which will be
specified later).
\end{enumerate}
\item[\textbf{Step 2)}] \underline{Delivery phase}
\begin{enumerate}
\item By the time the target device receives the requesting message,
the requesting device has moved to another position $C^{4}$ with the
mobility trace based on the RWMM. As illustrated in Fig.
\ref{2.1}, the desired content object delivered by the target device
chases the requesting device by executing essentially the same
procedure as the first delivery phase. \vspace{0.15cm} \item By
the time the FAP receives the requesting message, the requesting
device has moved to another position $C^{3}$. As illustrated in Fig.
\ref{2.2}, the desired content object delivered by the FAP is
delivered to a relay device in routing cell $C^{2}$ along another
straight line toward the requesting device within one time slot.
Thereafter, the relay chases the requesting device via multihop
until reaching the routing cell containing the requesting device. 
\end{enumerate}
\end{enumerate}
\par 
The overall procedure of the proposed content delivery routing protocol is summarized in Algorithm \ref{Alg1}.

\begin{algorithm}[t!]
\caption{The proposed content delivery routing protocol}\label{Alg1}
\begin{algorithmic}[1]
\STATE Step 1. First phase (requesting phase)
\bindent
\STATE Step 1-1. Device-to-device transmission
\IF {the closest holder is an FAP}
\STATE Step 1-2. Device-to-FAP transmission
\ENDIF
\eindent
\STATE Step 2: Second phase (delivery phase)
\bindent
\IF {the closest holder is an FAP}
\STATE Step 2-1. FAP-to-device transmission
\ENDIF
\STATE Step 2-2. Device-to-device transmission
\eindent
\end{algorithmic}
\end{algorithm}

\section{Throughput--Delay Trade-off}

In this section, we characterize a fundamental throughput--delay trade-off in terms of scaling laws for the content-centric mobile hybrid IoT network using the proposed content delivery routing. 
As stated in Section~\ref{routing}, we consider the nearest
neighbor multihop transmission between a requesting device and its
closest holder of the desired content object, where their distance
is crucially determined by the total number of replicas of content
object $m \in \mathcal{M}$, $A_m+B_m$. When replicas of each
content object are independently and uniformly distributed in the
caching phase, the average Euclidean distance from a requesting
device to its closest holder was shown to scale as the reciprocal of
the square root of the total number of holders of the desired
content object in the network (refer to~\cite{c1,a1} for more details). By applying this
argument to our network framework, we establish the
following lemma, which is essential to characterize the
throughput--delay trade-off.
\begin{lemma} \label{dist}
For any mobile device requesting content object $m \in \mathcal{M}$,
the average initial distance between any requesting device and its
closest holder of content object $m$ is
$\Theta\left(\frac{1}{\sqrt{A_m+B_m}}\right)$, where $A_m$ and
$B_m$ are the number of replicas of content object $m$ stored at
devices and FAPs, respectively.
\end{lemma}
\begin{IEEEproof}
The detailed proof of this argument is omitted here since it
basically follows the same line as the proof of~\cite[Lemma
3]{a1} with a slight modification.
\end{IEEEproof}

We are now ready to show our first main result.

\begin{thm} \label{tradeoff}
Suppose that the content delivery routing in Section \ref{routing}
is used for the content-centric mobile hybrid IoT network. Then, the
throughput--delay trade-off is given by

\begin{equation}
\label{equa6}
\lambda(n)  = \Theta\left( \frac{D(n)}{n \left(\sum^{M}_{m=1} \frac{p_m}{\sqrt{A_m+B_m}}\right)^2}\right)
\end{equation} whp, where
\begin{equation}
\label{equa91}
\lambda(n)=O\left(\frac{1}{\sum^{M}_{m=1} p_m \sqrt{\frac{n\log n}{A_m+B_m}}}\right) \nonumber
\end{equation}
and $p_m$ is the request probability of content object $m \in \mathcal{M}$.
\end{thm}

\begin{IEEEproof}
We compute the initial distance
of a randomly selected S--D pair into our cache-enabled network
setting. The length of the routing path of a requesting
message or a desired content object is shown to be determined by
the initial distance between a requesting device and its closest
holder of the desired content object, which is given by
$\Theta\left(\frac{1}{\sqrt{A_m+B_m}}\right)$ from Lemma
\ref{dist}.
As a result, the total number of hops along the routing paths of both the requesting message and desired content object scales as $\Theta\left(\frac{1}{\sqrt{a(n)(A_m+B_m})}\right)$. Since in our network, the average delay $D(n)$ is determined by the time taken from the moment that the requesting message leaves until the desired content object arrives at the requesting device, we have

\begin{equation}
\label{equa82}
D(n)=\Theta\left(\sum^{M}_{m=1}   \frac{p_m}{\sqrt{a(n)(A_m+B_m})}\right),
\end{equation}
where $p_m$ is the probability that each device requests content object $m \in \mathcal{M}$.\par
Similarly as in \cite{c1}, from the fact that the number of content objects passing through an arbitrary routing cell in each time slot is given by
 $O\left(n\sum^{M}_{m=1} p_m  \sqrt{\frac{a(n)}{A_m+B_m}}\right)$ whp,
the average per-device throughput is given by
\begin{equation}
\label{equa81}
\lambda(n)= \Theta\left(\frac{1}{n\sum^{M}_{m=1} p_m  \sqrt{\frac{a(n)}{A_m+B_m}}}\right)\text{whp},
\end{equation}
which is maximized when $a(n)=\Theta\left(\frac{\log n}{n}\right)$. Hence, using (\ref{equa82}) and (\ref{equa81}) leads to (\ref{equa6}), which completes the proof of the theorem.
\end{IEEEproof}

Theorem \ref{tradeoff} implies that the throughput--delay
trade-off is influenced by the total number of replicas of each
content object $m$, i.e., $A_m+B_m$. Due to the caching
constraints in (\ref{equa1}) and (\ref{equa01}), it is not
straightforward how to optimally allocate the sets of replicas,
${\lbrace A_m \rbrace}^M_{m=1}$ and ${\lbrace B_m
\rbrace}^M_{m=1}$, to show a net improvement in the overall
throughput--delay trade-off. In the next section, we introduce the
order-optimal cache allocation strategy to characterize the order-optimal
throughput--delay trade-off.

\begin{remark}
Based on the above result, it is not difficult to show that if
content delivery is performed via multihop, then the
throughput--delay trade-off for \textit{mobile} IoT networks is
identical to that for \textit{static} IoT networks
in~\cite{a1}. That is, we may conclude that performance on
the throughput--delay trade-off of both cache-enabled networks is the
same as far as the content delivery multihop routing protocols are
employed.
\end{remark}

According to the same analysis in~\cite{a1}, it is not difficult to show that when $\alpha\geq 3/2$, the order-optimal
throughput--delay trade-off in Theorem \ref{tradeoff} is
given by $\lambda(n)=\Theta\left(D(n)\right)$ by using device-to-device multihop
communication without FAP support. Therefore, for the rest of this
paper, we focus on the case where $\alpha < 3/2$ in solving the
optimal content replication problem.

\section{Order-Optimal Cache Allocation Strategy in Mobile Hybrid IoT Networks} \label{joint}

In this section, we characterize the order-optimal throughput--delay
trade-off of the content-centric mobile hybrid IoT network by
optimally selecting the replication sets ${\lbrace A_m
\rbrace}^M_{m=1}$ and ${\lbrace B_m \rbrace}^M_{m=1}$ in terms of scaling laws. We first
introduce our problem formulation in terms of maximizing the
throughput--delay trade-off. Then, we propose a content
replication strategy that jointly finds the number of replicas
cached at mobile devices and static FAPs, thus leading to the order-optimal
throughput--delay trade-off of our
network. Finally, We validate our analysis by numerically showing
our optimal solution to the cache allocation problem.

\subsection{Problem Formulation}
From Theorem \ref{tradeoff}, it can be shown that maximizing the throughput--delay trade-off is equivalent to maximizing the throughput $\lambda(n)$ for given delay $D(n)$. Thus, from the caching constraints in (\ref{equa1}) and (\ref{equa01}), we formulate the following optimization problem:
\begin{subequations}
\label{equa100}
\begin{align}
   \underset{{{\lbrace{A_m\rbrace}}^M_{m=1},{\lbrace B_m \rbrace}^M_{m=1}}}{\max}
        & \quad \lambda(n)  \label{cond11}\\
    \text{subject to}
        & \quad \sum_{m=1}^{M} A_m \leq nK_n  \,, \label{a}\\
        & \quad \sum_{m=1}^{M} B_m \leq f(n)K_{FAP} \,, \label{b}\\
        & \quad A_m \leq n  \text{ for }  m \in \mathcal{M} \,, \label{c}\\
        & \quad B_m \leq f(n)   \text{ for }  m \in \mathcal{M} \,, \label{d}\\
        & \quad A_m + B_m  \geq 1  \text{ for }  m \in \mathcal{M} \,.\label{e}
\end{align}
\end{subequations}\par
Since maximizing $\lambda(n)$ for given $D(n)$ is equivalent to minimizing the term $n\left(\sum_{m=1}^M \frac{p_m}{\sqrt{A_m+B+m}}\right)$ in (\ref{equa6}), the original problem in (\ref{equa100}) can be rewritten as
\begin{subequations}
\label{equa10}
\begin{align}
    \min_{{\lbrace{A_m\rbrace}}^M_{m=1},{\lbrace B_m \rbrace}^M_{m=1}}
        & \quad \sum_{m=1}^{M} \frac{p_m}{\sqrt{A_m+B_m}}  \label{cond1}\\
    \text{subject to} & \quad \text{(\ref{a})--(\ref{e}).} \label{f}
\end{align}
\end{subequations}\par
From the fact that the second derivatives of the objective
function~(\ref{cond1}) with respect to $A_m$ and $B_m$ for
$\forall m \in \mathcal{M}$ are non-negative, it is
straightforward to preserve the convexity of the objective
function. Note that the numbers of replicas of content
object $m$ stored at mobile devices and static FAPs, corresponding to $A_m$ and
$B_m$, respectively, are integer variables, which makes the
optimization problem non-convex and thus intractable. However, as
long as scaling laws are concerned in this work, the discrete
variables $A_m$ and $B_m$ for $m \in \mathcal{M}$ can be relaxed
to real numbers in $[1,\infty)$ so that the objective function in
(\ref{cond1}) becomes convex and differentiable. In addition,
since all inequality constraints (\ref{a})--(\ref{e}) are linear
functions, the problem in (\ref{equa10}) can be a convex
optimization problem.

\subsection{Analytical Results} \label{results}

Since the objective function is convex, we are able to use the
Lagrangian relaxation method for solving the problem in
(\ref{equa10}). In our work, we apply a novel \textit{variable
decoupling} technique for the replication sets ${\lbrace A_m
\rbrace}^M_{m=1}$ and ${\lbrace B_m \rbrace}^M_{m=1}$, which leads
to a much simpler analysis without fundamentally loosing order optimality,
compared to tackling the original problem that may not provide a
tractable closed-form solution. Notice that the total number of
replicas of content object $m \in \mathcal{M}$, $A_m+B_m$, can be
simplified as $\Theta(A_m)$ or $\Theta(B_m)$ according to the
relative size of $A_m$ and $B_m$. For analytical convenience,
given the optimal solution ${\lbrace A_m^* \rbrace}^M_{m=1}$ and
${\lbrace B_m^* \rbrace}^M_{m=1}$, let us define two subsets
$\mathcal{M}_1$ and $\mathcal{M}_2$ as the sets of content objects
such that $A_m^*+B_m^*=\Theta(A_m^*)$ and
$A_m^*+B_m^*=\Omega(f(n))$, respectively, which will be specified
in Theorem 2. We also define the subset $\mathcal{M}_3$ as the set
of content objects such that $A_m^*+B_m^*=\Theta(B_m^*)$. Note
that for $m \in \mathcal{M}_3$, $A_m=O(B_m)=O\left(f(n)\right)$.
Then, the optimization problem in (\ref{equa10}) is divided into
the following two optimization problems:
\begin{subequations}\label{equa1100}
\begin{align}
    \min_{{\lbrace{A_m\rbrace}}_{m\in \mathcal{M}_1}}
        & \sum_{m\in\mathcal{M}_1} \frac{p_m}{\sqrt{A_m}} \\
    \text{subject to}
        & \sum_{m=1}^{M} A_m \leq nK_n \label{Ac},  \\
        & A_m \leq n  \text{ for }  m \in \mathcal{M}_1 \,
\end{align}
\end{subequations}
and
\begin{subequations}\label{equa110}
\begin{align}
    \min_{{\lbrace{B_m\rbrace}}_{m\in\mathcal{M}_3}}
        & \sum_{m\in\mathcal{M}_3} \frac{p_m}{\sqrt{B_m}} \\
    \text{subject to}
        & \sum_{m=1}^M B_m \leq f(n)K_{FAP}, \label{Bc} \\
        & B_m \leq f(n)  \text{ for }  m \in \mathcal{M}_3 \,.
\end{align}
\end{subequations}
The Lagrangian function corresponding to (\ref{equa1100}) is given by
\begin{align}
\label{equa13}
&\mathcal{L}_1 \left( \lbrace A_m\rbrace _{m \in \mathcal{M}_1},\lambda,\lbrace w_m\rbrace _{m \in \mathcal{M}_1} \right)\nonumber\\&=\!\sum_{m\!\in\!\mathcal{M}_1}\! \frac{p_m}{\sqrt{\!A_m\!}}\!+\!\lambda\!\left(\!\sum_{m=1}^M \! A_m\!-\!n\!K_n\!\right)\!+\!\sum_{m\in\mathcal{M}_1} w_m(A_m-n),
\end{align}
where $w_m$, $\lambda$ $\in$ $\mathbb{R}$. The Karush--Kuhn--Tucker (KKT) conditions for (\ref{equa1100}) are then given by
\begin{align}
\frac{\partial \mathcal{L}_1 \left( \lbrace A_m^*\rbrace _{m \in \mathcal{M}_1},\lambda^*,\lbrace w_m^*\rbrace _{m \in \mathcal{M}_1} \right)}{\partial A_m^*}  =0& \label{AAA}\\
\lambda^*  \geq 0 &\nonumber\\
   w^*_m  \geq 0& \nonumber\\
   w^*_m(A_m^*-n)  =0& \label{A}\\
          \lambda^*\left(\sum_{m=1}^M A_m^*-nK_n\right)  =0\label{AA}
  \end{align}
for $m \in \mathcal{M}_1$.
Similarly, the Lagrangian function corresponding to (\ref{equa110}) is
\begin{equation}
\label{equa13b}
\begin{split}
&\mathcal{L}_2 \left( \lbrace B_m\rbrace _{m \in
\mathcal{M}_3},\mu,\lbrace \nu_m\rbrace _{m \in \mathcal{M}_3}
\right)\\&=\!\sum_{m\!\in\!\mathcal{M}_3}\!
\frac{p_m}{\sqrt{\!B_m\!}}\!+\!\mu\!\left(\!\sum_{m=1}^M\!
B_m\!-\!f(\!n\!)\!K_{F\!A\!P}\!\right)\!+\!\!\sum_{m\!\in\!\mathcal{M}_3}\!\!
\nu_m(\!B_m\!-\!f(\!n\!))\!,
\end{split}
\end{equation}
where $\nu_m$, $\mu$ $\in$ $\mathbb{R}$. Then for $\forall m \in \mathcal{M}_3$, the KKT conditions for (\ref{equa110}) state that
\begin{align}
\frac{\partial \mathcal{L}_2 \left( \lbrace B_m^*\rbrace _{m \in \mathcal{M}_3},\mu^*,\lbrace \nu_m^*\rbrace _{m \in \mathcal{M}_3} \right) }{\partial B_m^*} & =0 \nonumber \\
\mu^* & \geq 0  \nonumber\\
   \nu^*_m & \geq 0 \nonumber\\
   \nu^*_m(B_m^*-f(n))&  =0  \nonumber\\
          \mu^*\left(\sum_{m=1}^M B_m^*-f(n)K_{FAP}\right)  &=0. \label{BBB}
  \end{align}

We start from introducing the following lemma, which plays an important role in solving our content replication problem.

\begin{lemma} \label{m2}
Suppose that $\alpha < 3/2$ and the content delivery
routing in Section \ref{routing} is used for the content-centric
mobile hybrid IoT network. Then, the optimal solution to
\eqref{equa1100}, denoted by $A_m^*$, is non-increasing with $m
\in \mathcal{M}_1$ and the optimal solution to \eqref{equa110},
denoted by $B_m^*$, is non-increasing with $m \in \mathcal{M}_3$.
\end{lemma}

\begin{IEEEproof}
Refer to Appendix~\ref{PF:Lemma2}.
\end{IEEEproof}

From the above lemma, the following important theorem can be
established, which shows the optimal total number of replicas of
contents $m \in \mathcal{M}$ at both mobile devices and static FAPs in terms of scaling laws.

\begin{thm} \label{Thm:Theorem2}
Suppose that $\alpha < 3/2$ and the content delivery routing in
Section \ref{routing} is used for the content-centric mobile hybrid IoT network. If $\alpha \leq
\frac{3(\gamma-\beta)}{2(\delta+\gamma-1)}$, then the order-optimal
solution to (\ref{equa10}) is given by
\begin{equation}
\label{equa12}
A^{*}_m+B^{*}_m=
 \Theta\left(m^{-\frac{2\alpha}{3}}n^{\beta+\delta-\gamma\left(1-\frac{2\alpha}{3}\right)}\right).\nonumber
\end{equation}
If $\frac{3(\gamma-\beta)}{2(\delta+\gamma-1)}< \alpha<\frac{3}{2}$, then it is given by
\begin{equation}
\label{equa11}A^{*}_m\!+\!B^{*}_m\!=\!
  \begin{cases}
    \!\Theta \!\left(\!m^{-\frac{2\alpha}{3}}n^{\delta+(1-\delta)\frac{2\alpha}{3}}\!\right) & \textnormal{\!\!for } m \!\in \!\mathcal{M}_1, \\
   \!\Theta\!\left(n^\delta \right) & \textnormal{\!\!for } m\!\in\!\mathcal{M}_2\setminus\mathcal{M}_1, \\
   \!\Theta\!\left(\!m^{-\frac{2\alpha}{3}}n^{\beta+\delta-\gamma\left(1-\frac{2\alpha}{3}\right)}\!\right) & \textnormal{\!\!for } m \!\in\!\mathcal{M}\setminus\mathcal{M}_2,
  \end{cases}\nonumber
\end{equation}
where $\mathcal{M}_1 = \lbrace 1,...,m_1-1 \rbrace$ and $\mathcal{M}_2 = \lbrace 1,...,m_2-1 \rbrace$. Here, $m_1=\Theta\left(n^{1-\delta}\right)$ and $m_2=\Theta\left(n^{\gamma-(\gamma-\beta)\frac{3}{2\alpha}}\right)$.
\end{thm}
\begin{IEEEproof}
Refer to Appendix~\ref{PF:Theorem2}.
\end{IEEEproof}
\begin{figure}[t!]
    \centering
     \subfigure[$\alpha \leq \frac{3(\gamma-\beta)}{2(\delta+\gamma-1)}$]{\leavevmode
  \begin{tikzpicture}[scale=0.46]
\draw[->] (0,0) -- (10,0) node[anchor=north] {{\small $m$}};
\draw   (0,0) node[anchor=north] {0};
\draw[->] (0,0) -- (0,8) node[anchor=east] {{\small $A_m^*\!+\!B_m^*$}};
\draw[blue] (0.1,7) --(0.3,3.85);
\draw[blue](0.3,3.85) to [bend right] (1.8,1.8);
\draw[blue] (1.8,1.8) parabola [bend at end] (9.5,1);
\end{tikzpicture}
  \leavevmode \epsfxsize=0.22\textwidth
       \label{5a.1}
}
       \subfigure[$\frac{3(\gamma-\beta)}{2(\delta+\gamma-1)}<\alpha <  \frac{3}{2}$]{\leavevmode
  \begin{tikzpicture}[scale=0.46]
\draw[->] (0,0) -- (10,0) node[anchor=north] {{\small $m$}};
\draw   (0,0) node[anchor=north] {0};
\draw[->] (0,0) -- (0,8) node[anchor=east] {{\small $A_m^*\!+\!B_m^*$}};
\draw[blue] (0.1,7) parabola [bend at end](0.5,2);
\draw[blue](0.5,2) -- (1.8,2);
\draw[blue] (1.8,2) parabola [bend at end] (9.5,0.6);
\draw(7,3) node[anchor=east] {{\small $A_m^*\!+\!B_m^*\!=\!\Theta\!\left(\!f(n)\!\right)$}};
\draw[dashed,red] (1.1,2) ellipse (1.1cm and 0.25cm);
\end{tikzpicture}
 \leavevmode \epsfxsize=0.22\textwidth
        \label{5a.2}
 }
    \caption{\small{The order-optimal cache allocation strategy with respect to the content object $m$.}}
    \label{5a}
\end{figure}
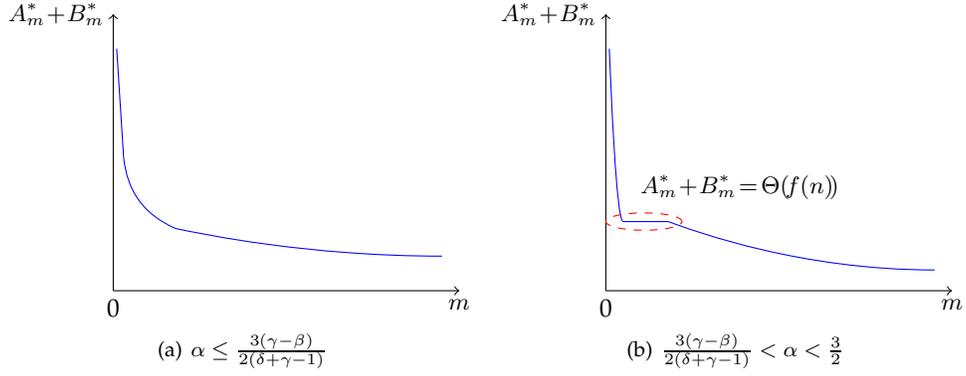
The order-optimal solution in Theorem \ref{Thm:Theorem2} is illustrated in Fig. \ref{5a}. For $\alpha \leq \frac{3(\gamma-\beta)}{2(\delta+\gamma-1)}$, the optimal number of replicas of content object $m$, $A_{m}^*+B_m^*$, is shown to monotonically decrease with $m$. For $\frac{3(\gamma-\beta)}{2(\delta+\gamma-1)}< \alpha<\frac{3}{2}$, there exists a set of content objects such that $A_{m}^*+B_m^*$ scales as $\Theta(f(n))$. \par
Now, we turn to describing our replication strategy by individually choosing the replication sets ${\lbrace A_m^* \rbrace}^M_{m=1}$ and ${\lbrace B_m^* \rbrace}^M_{m=1}$. We introduce the following proposition, which exhibits that the resulting ${\lbrace A_m^* \rbrace}^M_{m=1}$ and ${\lbrace B_m^* \rbrace}^M_{m=1}$ still guarantee the order optimality as long as scaling laws are concerned.
\begin{proposition} \label{cache1}
Suppose that $\alpha<\frac{3}{2}$ and the content delivery routing
in Section~\ref{routing} is used for the content-centric mobile hybrid IoT network. Then, the order-optimal individual replication sets
${\lbrace A_m^* \rbrace}^M_{m=1}$ and ${\lbrace B_m^*
\rbrace}^M_{m=1}$ are given by
\begin{equation}\label{M1}
A^{*}_m\!\!=\!\!
  \begin{cases}
       \!\!\Theta\!\left(\!m^{\!-\!\frac{2\!\alpha}{3}}\!n^{\!\min\!\left\lbrace\!\beta+\!\delta\!-\!\gamma\!(\!1\!-\!\frac{2\!\alpha}{3})\!,\delta\!+\!(\!1\!-\!\delta\!)\!\frac{2\!\alpha}{3}\!\right\rbrace}\!\right) & \!\!\textnormal{for\! } m \!\in\!\mathcal{M}\!_1\!\cap\!\mathcal{M}\!_2, \\
   \!0 & \!\!\textnormal{for\! } m \!\in\! \mathcal{M}\!\!\setminus\!\!\left(\!\mathcal{M}\!_1\!\cap\!\mathcal{M}\!_2\!\right)
  \end{cases}
\end{equation}
and
\begin{equation}
\hspace{-0.65cm}
   B^{*}_m\!=\!
  \begin{cases}
      \!\!\Theta\!\left(n^\delta \right) & \textnormal{ for } m \in\mathcal{M}_2,  \\
   \!\!\Theta\!\left(m^{-\frac{2\alpha}{3}}n^{\beta+\delta-\gamma\left(1-\frac{2\alpha}{3}\right)}\right) & \textnormal{ for } m \in \mathcal{M}\setminus\mathcal{M}_2,
  \end{cases}\nonumber
\end{equation}
respectively, where $\mathcal{M}_1 = \lbrace 1,...,m_1-1 \rbrace$ and $\mathcal{M}_2 = \lbrace 1,...,m_2-1 \rbrace$. Here, $m_1=\Theta\left(n^{1-\delta}\right)$ and $m_2=\Theta\left(n^{\gamma-(\gamma-\beta)\frac{3}{2\alpha}}\right)$.
\end{proposition}
\begin{IEEEproof}
We prove the proposition by individually selecting ${\lbrace A_m^* \rbrace}^M_{m=1}$ and ${\lbrace B_m^* \rbrace}^M_{m=1}$ that satisfy the order-optimal solution ${\lbrace A_m^*+B_m^* \rbrace}^M_{m=1}$ in Theorem \ref{Thm:Theorem2} according to the following two cases depending on the value of $\alpha$.\par
Let us first focus on the case where $\alpha \leq \frac{3(\gamma-\beta)}{2(\delta+\gamma-1)}$. For $m \in \mathcal{M}_2$ where $A_m^*+B_m^*=\Omega(f(n))\left(=\Omega\left(n^\delta\right)\right)$, we set $A_m^*=\Theta\left(m^{-\frac{2\alpha}{3}}n^{\beta+\delta-\gamma\left(1-\frac{2\alpha}{3}\right)}\right)$ and $B_m^*=\Theta\left(n^\delta\right)$. For $\mathcal{M}\setminus\mathcal{M}_2$ where $A_m^*+B_m^*=o(f(n))$, we set $A_m^*=0$ and $B_m^*=\Theta\left(m^{-\frac{2\alpha}{3}}n^{\beta+\delta-\gamma\left(1-\frac{2\alpha}{3}\right)}\right)$.   \par
Next, we consider the case where $\frac{3(\gamma-\beta)}{2(\delta+\gamma-1)}< \alpha<\frac{3}{2}$. For $m \in \mathcal{M}_1$ where $A_m^*+B_m^*=\Theta\left(A_m^*\right)$, we set $A_m^*=\Theta \left(m^{-\frac{2\alpha}{3}}n^{\delta+(1-\delta)\frac{2\alpha}{3}}\right)$ and $B_m^*=\Theta\left(n^\delta\right)$. For $m \in \mathcal{M}_2\setminus \mathcal{M}_1$ where $A_m^*+B_m^*=\Theta(f(n))$, we set $A_m^*=0$ and $B_m^*=\Theta\left(n^\delta\right)$. As in the previous case, we also set $A_m^*=0$ and $B_m^*=\Theta\left(m^{-\frac{2\alpha}{3}}n^{\beta+\delta-\gamma\left(1-\frac{2\alpha}{3}\right)}\right)$ for $m \in \mathcal{M}\setminus \mathcal{M}_2$.  \par
From the fact that $m_1=\Theta\left(n^{1-\delta}\right)$ and $m_2=\Theta\left(n^{\gamma-(\gamma-\beta)\frac{3}{2\alpha}}\right)$, it follows that $\mathcal{M}_1\cap\mathcal{M}_2=\mathcal{M}_2$ if $\alpha \leq \frac{3(\gamma-\beta)}{2(\delta+\gamma-1)}$. Otherwise, $\mathcal{M}_1\cap\mathcal{M}_2=\mathcal{M}_1$. As a consequence, according to the above setting, $A_m^*$ can be written in a single expression in (\ref{M1}). This completes the proof of the proposition.
\end{IEEEproof}
\begin{figure}[t!]
    \centering
     \subfigure[$\alpha \leq \frac{3(\gamma-\beta)}{2(\delta+\gamma-1)}$]{\leavevmode
  \begin{tikzpicture}[scale=0.46]
\draw[->] (0,0) -- (10,0) node[anchor=north] {{\small $m$}};
\draw   (0,0) node[anchor=north] {0};
\draw[->] (0,0) -- (0,8) node[anchor=east] {{\small $A_m^*,B_m^*$}};
\draw[blue] (0.1,7) --(0.3,3.85);
\draw[blue] (0.3,3.85) --(0.3,0.08);
\draw[blue](0.3,0.08) -- (9.5,0.08);
\draw[blue](0,3.85)-- (0.3,3.85);
\draw[blue](0.3,3.85) to [bend right] (1.8,1.8);
\draw[blue] (1.8,1.8) parabola [bend at end] (9.5,1);
\draw   (6,4.5) node[anchor=east] {{ $B_m^*=\Theta\left(f(n)\right)$}};
\draw[dashed,red] (0.2,3.85) ellipse (0.32cm and 0.2cm);
\draw   (3,6.5) node[anchor=east] {{ $A_m^*$}};
    \draw [black, ->          ] (1.5,6) -- (0.25,5.8);
    \draw   (6,2.5) node[anchor=east] {{ $B_m^*$}};
    \draw [black, ->          ] (5,1.8) -- (6,1.3);
\end{tikzpicture}
  \leavevmode \epsfxsize=0.22\textwidth
       \label{6a.1}
}
       \subfigure[$\frac{3(\gamma-\beta)}{2(\delta+\gamma-1)}<\alpha <  \frac{3}{2}$]{\leavevmode
  \begin{tikzpicture}[scale=0.46]
\draw[->] (0,0) -- (10,0) node[anchor=north] {{ $m$}};
\draw   (0,0) node[anchor=north] {0};
\draw[->] (0,0) -- (0,8) node[anchor=east] {{ $A_m^*,B_m^*$}};
\draw[blue] (0.1,7) parabola [bend at end](0.5,2);
\draw[blue](0.5,2) -- (0.5,0.08);
\draw[blue](0.5,0.08) -- (9.5,0.08);
\draw[blue](0,2) -- (1.8,2);
\draw[blue] (1.8,2) parabola [bend at end] (9.5,0.6);
\draw   (6,3.1) node[anchor=east] {{ $B_m^*=\Theta\left(f(n)\right)$}};
\draw[dashed,red] (1.1,2) ellipse (1.1cm and 0.25cm);
\draw   (3,7) node[anchor=east] {{ $A_m^*$}};
    \draw [black, ->          ] (1.75,6.5) -- (0.25,5.8);
    \draw   (6.8,2) node[anchor=east] {{ $B_m^*$}};
    \draw [black, ->          ] (5.5,1.5) -- (4.3,1.3);
\end{tikzpicture}
  \leavevmode \epsfxsize=0.22\textwidth
        \label{6a.2}
 }
 \caption{\small{The replication sets ${\lbrace A_m^* \rbrace}^M_{m=1}$ and ${\lbrace B_m^* \rbrace}^M_{m=1}$ with respect to the content object $m$.}}
    \label{6aa}
\end{figure}
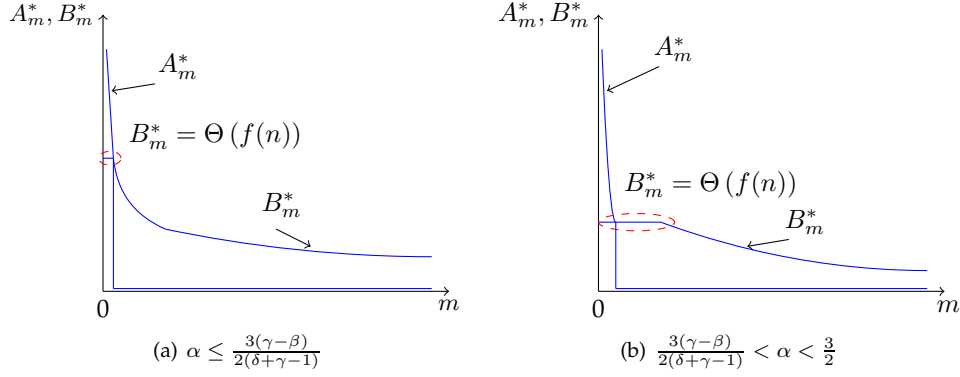
The order-optimal replication strategy in Proposition~\ref{cache1} is
illustrated in Fig.~\ref{6aa}. From this result, the following
insightful observations are made.
\begin{remark} \label{re1}
From Fig.~\ref{6aa}, it is observed that highly popular content
objects, whose number of replicas is $\omega\left(f(n)\right)$,
are mainly served by device-to-device multihop routing while the rest
of the content objects are served by deployed FAPs. More
specifically, from Proposition~\ref{cache1}, caching content
objects $m\in\mathcal{M}_1\cap\mathcal{M}_2$ mostly at mobile
devices is order-optimal in terms of throughput--delay trade-off. Thus,
this replication strategy sheds light on how to cache in
large-scale content-centric mobile hybrid IoT networks.
\end{remark}
From Theorems \ref{tradeoff} and \ref{Thm:Theorem2} and Proposition
\ref{cache1}, we characterize the order-optimal throughput--delay
trade-off. To be specific, the objective function in (\ref{cond1})
is given by
\begin{align}
\sum_{m=1}^{M} \frac{p_m}{\sqrt{A_m^*+B_m^*}}&=\Theta\left(\frac{n^{\gamma(\frac{3}{2}-\alpha)-\frac{\beta+\delta}{2}}}{H_{\alpha}(M)}\right), \nonumber
\end{align}
where $H_{\alpha}(M)=\sum_{i=1}^M i^{-\alpha}$. For $\frac{3(\gamma-\beta)}{2(\delta+\gamma-1)}<\alpha <  \frac{3}{2}$, we have
\begin{align}\label{equa34}
&\sum_{m=1}^{M} \frac{p_m}{\sqrt{A_m^*+B_m^*}}\nonumber\\&=\!\sum_{m=1}^{m_1-1}\! \frac{p_m}{\sqrt{A_m^*\!+\!B_m^*}}\!+\!\sum_{m=m_1}^{m_2-1}\! \frac{p_m}{\sqrt{A_m^*\!+\!B_m^*}}\!+\!\sum_{m=m_2}^{M}\! \frac{p_m}{\sqrt{A_m^*\!+\!B_m^*}} \nonumber \\&=
\Theta\!\left(\!\frac{n^{(1-\delta)(\frac{3}{2}-\alpha)-\frac{1}{2}}}{H_{\alpha}(M)}\!\right)\!+\!\Theta\!\left(\!\frac{\!n^{-\!\frac{\delta}{2}}\max\lbrace H_{\alpha}(m_1),H_{\alpha}(m_2)\rbrace}{H_{\alpha}(M)}\!\right)\nonumber\\&+\Theta\left(\frac{n^{\gamma(\frac{3}{2}-\alpha)-\frac{\beta+\delta}{2}}}{H_{\alpha}(M)}\right).
\end{align}
Let the first, second, and third terms in the right-hand side of (\ref{equa34}) be denoted by $F_1$, $F_2$, and $F_3$, respectively. Then, it is seen that $F_2=\Theta(F_1)$ for $1<\alpha<\frac{3}{2}$ and $F_2=O(F_3)$ for $\alpha\leq 1$. In addition, for $\alpha \leq\frac{3(\gamma-\beta)}{2(\delta+\gamma-1)}$, the objective function $\sum_{m=1}^{M} \frac{p_m}{\sqrt{A_m^*+B_m^*}}$ in (\ref{cond1}) is shown to scale as $F_3$. Thus, we need to compare the relative size of $F_1$ and $F_3$ according to the four scaling parameters $\alpha$, $\delta$, $\beta$, and $\gamma$ to scrutinize the scaling behavior of (\ref{cond1}). In our work, we partition the entire parameter space, including the case where $\alpha\geq \frac{3}{2}$, into three operating regimes as follows (refer to Fig. \ref{regime}).
\begin{itemize}
\item Regime I (High Zipf exponent regime): $\lbrace
\alpha\vert\alpha \geq \frac{3}{2}\rbrace$ \item Regime II (Medium
Zipf exponent regime): $ \left\lbrace \alpha \bigg\vert
1+\frac{\gamma-\beta}{2(\gamma+\delta-1)}\leq\alpha < \frac{3}{2}
\right\rbrace  \nonumber $ \item Regime III (Low Zipf exponent
regime): $ \left\lbrace \alpha \bigg\vert \alpha <
1+\frac{\gamma-\beta}{2(\gamma+\delta-1)} \right\rbrace \nonumber
$
\end{itemize}

We are now ready to characterize the order-optimal throughput--delay trade-off in the following theorem.

\begin{thm} \label{thm2} Suppose that the content delivery routing in Section \ref{routing} and the order-optimal replication strategy in Proposition \ref{cache1} are used for the content-centric mobile hybrid IoT network. Then, according to the Zipf exponent $\alpha$ and the scaling parameters $\gamma$, $\delta$, and $\beta$, the order-optimal throughput--delay trade-off is given by
\begin{equation}
\lambda(n)=\Theta\left(\frac{D(n)}{n^{b}}\right),  \textnormal{ where }  \lambda(n)=O\left(\frac{1}{\sqrt{n^{b+\epsilon}}}\right),\nonumber
\end{equation}
for an arbitrarily small constant $\epsilon> 0$. Here,
\begin{equation}
b=
\begin{cases}
0  & \textnormal{ in Regime I},\\
(1-\delta)(3-2\alpha) &  \textnormal{ in Regime II},\\
1-\delta-\beta+\min \left\lbrace3-2\alpha,1\right\rbrace  \gamma & \textnormal{ in Regime III}.
\end{cases}\nonumber
\end{equation}
\end{thm}
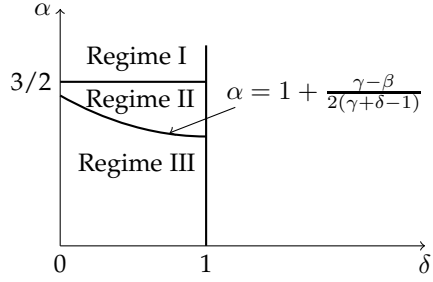
\begin{figure}[t!]
    \centering
     \begin{tikzpicture}[scale=0.485]
\draw[->] (0,0) -- (10,0) node[anchor=north] {$\delta$};
\draw   (0,0) node[anchor=north] {0};
\draw   (4,0) node[anchor=north] {1};
\draw   (0,4.5) node[anchor=east] {3/2};
\draw   (10.2,4.17) node[anchor=east] {$\alpha=1+\frac{\gamma-\beta}{2(\gamma+\delta-1)}$};
\draw   (2,5.125) node{{ Regime I}};

\draw   (2.1,3.95) node{{ Regime II}};

\draw   (2,2.2) node{{  Regime III}};

\draw[->] (0,0) -- (0,6.5) node[anchor=east] {$\alpha$};
\draw[thick] (4,0) -- (4,5.5);
\draw[thick] (0,4.5) -- (4,4.5);
\draw[thick] (0,4.125) parabola[bend at end] (4,3);
\draw [black, ->          ] (4.8,3.8) -- (3,3.1);
\end{tikzpicture}
    \caption{\small{The  operating  regimes  on  the  throughput--delay trade-off  with  respect to $\alpha$, $\delta$, $\beta$, and $\gamma$.}}
    \label{regime}
\end{figure}
\begin{IEEEproof}
In Regime I, $\lambda(n)=\Theta\left(D(n)\right)$ is achieved. In
Regime II, it is shown that (\ref{cond1}) scales as
$F_1=\Theta\left(n^{(1-\delta)(\frac{3}{2}-\alpha)-\frac{1}{2}}\right)$,
thereby resulting in $b=(1-\delta)(3-2\alpha)$ from (\ref{equa6}).
Now, let us turn to Regime III as follows: for
$1<\alpha<1+\frac{\gamma-\beta}{2(\gamma+\delta-1)}$,
(\ref{cond1}) scales as
$F_3=\Theta\left(n^{\gamma(\frac{3}{2}-\alpha)-\frac{\delta+\beta}{2}}\right)$;
and for $\alpha\leq 1$, (\ref{cond1}) scales as
$F_3=\Theta\left(n^{\frac{\gamma}{2}-\frac{\delta+\beta}{2}}\right)$.
Thus, we have $b=1-\delta-\beta+\min
\left\lbrace3-2\alpha,1\right\rbrace$. This completes the proof of
the theorem.
\end{IEEEproof}
The impact and benefits of FAPs equipped with a finite-size cache
are explicitly addressed according to each operating regime on the
order-optimal throughput--delay trade-off.

\begin{remark}
In Regime I (i.e., the high Zipf exponent regime), the best
performance $\lambda(n)=\Theta\left(D(n)\right)$ is achieved by
using device-to-device multihop communication, and thus the use of
FAPs does not further improve the performance.
This is because highly popular contents objects
are dominant in the regime. On the other hand, in Regimes II and
III (i.e., the medium and low Zipf exponent regimes), it turns out
that the supplemental cache space $f(n)K_{FAP}$ in the mobile hybrid IoT network significantly improves
the network performance over the mobile IoT network with no FAP. Interestingly, in Regime II, the order-optimal trade-off is
shown to be the same as in the content-centric \textit{static} IoT network case assuming FAPs
equipped with the infinite-size cache (or equivalently, the
infinite-speed backhaul-aided cache)~\cite{a1}.
\end{remark}

\section{Numerical Evaluation}
\begin{table}[t!]
\centering \caption{The simulation enviroment}\label{table1}
\resizebox{0.425\textwidth}{!}{
\begin{tabular}
{ |c|c|c| }  \hline
 Symbol & Description & Value\\
\hline\hline
  $n$ & The number of mobile devices & 300\\
  \hline
  $M$ & The size of the library & 200\\
    \hline
  $f(n)$ & The number of FAPs & 50 \\
   \hline
  $K_{FAP}$ & The cache size of each FAP & 50 \\
    \hline
  $K_n$ & The cache size of each device & 2\\  \hline
  \end{tabular}
 }
\end{table}

In this section, to validate the analytical results shown in
Section \ref{results}, we perform intensive numerical evaluation
by numerically solving the optimization problem in (\ref{equa10})
according to finite values of the system parameters $n$, $M$,
$K_n$, $K_{FAP}$, and $f(n)$ in Table \ref{table1}, where the
exponents $\gamma$, $\beta$, and $\delta$ are given by $0.93$,
$0.69$, and $0.69$, respectively. As in Section \ref{results}, we
consider two cases $\alpha \leq
\frac{3(\gamma-\beta)}{2(\delta+\gamma-1)}$ and
$\frac{3(\gamma-\beta)}{2(\delta+\gamma-1)}< \alpha<\frac{3}{2}$,
where $\frac{3(\gamma-\beta)}{2(\delta+\gamma-1)}=0.59$. To
account for a distinct feature of the optimal solution to
(\ref{equa10}) in each case, the following two values of $\alpha$
are used: $\alpha =0.55$  and $\alpha =1.2$.

\begin{figure}[t!]
    \centering
     \subfigure[$\alpha=0.55$]{\leavevmode
  \begin{tikzpicture}[scale=0.65]
     \begin{axis}[
        ylabel={\LARGE $A_m^*+B_m^*$},
        xlabel={\LARGE $m$},xmin=0, xmax=200,ymin=0, ymax=100]
       \addplot[smooth,blue, thick] plot coordinates {
        (1,98.21)
        (2,76.14)
        (3,65.63)
        (4,59.07)
        (5,54.43)
        (6,50.91)
        (7,48.12)
        (8,45.82)
        (9,43.89)
        (10,42.22)
        (11,40.77)
        (12,39.5)
        (13,38.36)
        (14,37.33)
        (15,36.4)
        (16,35.55)
        (17,34.77)
        (18,34.05)
        (19,33.37)
        (20,32.76)
        (21,32.18)
        (22,31.63)
        (23,31.11)
        (24,30.64)
        (25,30.18)
        (26,29.75)
        (27,29.34)
        (28,28.95)
        (29,28.58)
        (30,28.23)
        (31,27.89)
        (32,27.57)
        (33,27.26)
        (34,26.96)
        (35,26.67)
        (36,26.4)
        (37,26.14)
        (38,25.89)
        (39,25.64)
        (40,25.41)
        (41,25.17)
        (42,24.95)
        (43,24.74)
        (44,24.53)
        (45,24.33)
        (46,24.14)
        (47,23.94)
        (48,23.76)
        (49,23.58)
        (50,23.41)
        (51,23.24)
        (52,23.08)
        (53,22.91)
        (54,22.76)
        (55,22.61)
        (56,22.46)
        (57,22.31)
        (58,22.17)
        (59,22.03)
        (60,21.89)
        (61,21.76)
        (62,21.63)
        (63,21.51)
        (64,21.39)
        (65,21.26)
        (66,21.14)
        (67,21.03)
        (68,20.91)
        (69,20.8)
        (70,20.69)
        (71,20.58)
        (72,20.48)
        (73,20.37)
        (74,20.27)
        (75,20.18)
        (76,20.08)
        (77,19.98)
        (78,19.88)
        (79,19.79)
        (80,19.7)
        (81,19.61)
        (82,19.52)
        (83,19.43)
        (84,19.35)
        (85,19.27)
        (86,19.18)
        (87,19.11)
        (88,19.02)
        (89,18.94)
        (90,18.87)
        (91,18.8)
        (92,18.71)
        (93,18.64)
        (94,18.56)
        (95,18.5)
        (96,18.43)
        (97,18.36)
        (98,18.29)
        (99,18.23)
        (100,18.16)
        (101,18.09)
        (102,18.02)
        (103,17.96)
        (104,17.9)
        (105,17.83)
        (106,17.77)
        (107,17.71)
        (108,17.64)
        (109,17.58)
        (110,17.53)
        (111,17.48)
        (112,17.36)
        (113,17.41)
        (114,17.36)
        (115,17.25)
        (116,17.2)
        (117,17.1)
        (118,17.08)
        (119,17.03)
        (120,16.98)
        (121,16.93)
        (122,16.88)
        (123,16.83)
        (124,16.78)
        (125,16.73)
        (126,16.68)
        (127,16.63)
        (128,16.58)
        (129,16.54)
        (130,16.49)
        (131,16.44)
        (132,16.39)
        (133,16.35)
        (134,16.31)
        (135,16.26)
        (136,16.22)
        (137,16.26)
        (138,16.13)
        (139,16.09)
        (140,16.05)
        (141,16.01)
        (142,15.97)
        (143,15.93)
        (144,15.88)
        (145,15.84)
        (146,15.81)
        (147,15.76)
        (148,15.73)
        (149,15.68)
        (150,15.65)
        (151,15.61)
        (152,15.57)
        (153,15.53)
        (154,15.49)
        (155,15.46)
        (156,15.43)
        (157,15.39)
        (158,15.36)
        (159,15.32)
        (160,15.28)
        (161,15.24)
        (162,15.21)
        (163,15.18)
        (164,15.14)
        (165,15.11)
        (166,15.08)
        (167,15.04)
        (168,15.01)
        (169,14.97)
        (170,14.95)
        (171,14.92)
        (172,14.88)
        (173,14.85)
        (174,14.82)
        (175,14.79)
        (176,14.76)
        (177,14.72)
        (178,14.69)
        (179,14.67)
        (180,14.64)
        (181,14.6)
        (182,14.57)
        (183,14.55)
        (184,14.52)
        (185,14.49)
        (186,14.46)
        (187,14.43)
        (188,14.4)
        (189,14.38)
        (190,14.35)
        (191,14.32)
        (192,14.29)
        (193,14.26)
        (194,14.24)
        (195,14.21)
        (196,14.19)
        (197,14.15)
        (198,14.13)
        (199,14.11)
        (200,14.08)
            };
    \end{axis}
    \end{tikzpicture}
  \leavevmode \epsfxsize=0.22\textwidth
       \label{5.1}
}
       \subfigure[$\alpha=1.2$]{\leavevmode
  \begin{tikzpicture}[scale=0.65]
    \begin{axis}[
        ylabel={\LARGE $A_m^*+B_m^*$},
        xlabel={\LARGE $m$},xmin=0, xmax=200,ymin=0, ymax=300]
        \addplot[smooth,blue, thick] plot coordinates {
        (1,301.91)
        (2,173.94)
        (3,100.16)
        (4,83.88)
        (5,72.56)
        (6,64.19)
        (7,57.75)
        (8,52.74)
        (9,50.83)
        (10,50.34)
        (11,50.18)
        (12,50.09)
        (13,50.01)
        (14,49.92)
        (15,49.77)
        (16,49.54)
        (17,49.16)
        (18,48.53)
        (19,47.64)
        (20,46.51)
        (21,45.21)
        (22,43.84)
        (23,42.47)
        (24,41.13)
        (25,39.86)
        (26,38.66)
        (27,37.53)
        (28,36.47)
        (29,35.47)
        (30,34.53)
        (31,33.64)
        (32,32.79)
        (33,32)
        (34,31.25)
        (35,30.53)
        (36,29.86)
        (37,29.21)
        (38,28.59)
        (39,28.01)
        (40,27.44)
        (41,26.91)
        (42,26.4)
        (43,25.9)
        (44,25.43)
        (45,24.98)
        (46,24.54)
        (47,24.13)
        (48,23.72)
        (49,23.33)
        (50,22.96)
        (51,22.6)
        (52,22.25)
        (53,21.92)
        (54,21.59)
        (55,21.29)
        (56,20.98)
        (57,20.69)
        (58,20.4)
        (59,20.12)
        (60,19.86)
        (61,19.6)
        (62,19.34)
        (63,19.1)
        (64,18.86)
        (65,18.63)
        (66,18.4)
        (67,18.18)
        (68,17.79)
        (69,17.76)
        (70,17.55)
        (71,17.36)
        (72,17.16)
        (73,16.97)
        (74,16.79)
        (75,16.61)
        (76,16.44)
        (77,16.27)
        (78,16.1)
        (79,15.94)
        (80,15.78)
        (81,15.62)
        (82,15.47)
        (83,15.32)
        (84,15.17)
        (85,15.03)
        (86,14.89)
        (87,14.75)
        (89,14.49)
        (91,14.23)
        (92,14.11)
        (93,13.99)
        (94,13.87)
        (95,13.75)
        (96,13.64)
        (97,13.52)
        (98,13.41)
        (99,13.3)
        (100,13.2)
        (101,13.09)
        (102,12.99)
        (103,12.89)
        (104,12.79)
        (105,12.69)
        (106,12.6)
        (107,12.5)
        (108,12.41)
        (109,12.32)
        (110,12.23)
        (111,12.14)
        (112,12.05)
        (113,11.97)
        (114,11.88)
        (115,11.8)
        (116,11.72)
        (117,11.64)
        (118,11.56)
        (119,11.48)
        (120,11.41)
        (121,11.33)
        (122,11.26)
        (123,11.18)
        (124,11.11)
        (125,11.04)
        (126,10.97)
        (127,10.9)
        (128,10.83)
        (129,10.77)
        (130,10.7)
        (131,10.63)
        (132,10.57)
        (133,10.51)
        (134,10.44)
        (135,10.38)
        (136,10.32)
        (137,10.26)
        (138,10.2)
        (139,10.14)
        (140,10.08)
        (141,10.03)
        (142,9.97)
        (143,9.91)
        (144,9.86)
        (145,9.8)
        (146,9.75)
        (147,9.7)
        (148,9.64)
        (149,9.59)
        (150,9.54)
        (151,9.49)
        (152,9.44)
        (153,9.39)
        (154,9.34)
        (155,9.29)
        (156,9.25)
        (157,9.2)
        (158,9.15)
        (159,9.11)
        (160,9.06)
        (161,9.02)
        (162,8.97)
        (163,8.93)
        (164,8.88)
        (165,8.84)
        (166,8.88)
        (167,8.76)
        (168,8.71)
        (169,8.67)
        (170,8.63)
        (171,8.59)
        (172,8.55)
        (173,8.51)
        (174,8.47)
        (175,8.43)
        (176,8.4)
        (177,8.36)
        (178,8.32)
        (179,8.28)
        (180,8.24)
        (181,8.21)
        (182,8.17)
        (183,8.14)
        (184,8.1)
        (185,8.07)
        (186,8.03)
        (187,8)
        (188,7.96)
        (189,7.93)
        (190,7.9)
        (191,7.86)
        (192,7.83)
        (193,7.8)
        (194,7.77)
        (195,7.73)
        (196,7.7)
        (197,7.67)
        (198,7.64)
        (199,7.61)
        (200,7.58)
    };
           \draw    (120,75) node[anchor=east] {{\LARGE $A_m^*+B_m^*\approx 50$}};
    \draw[dashed,red] (15,50) ellipse (0.35cm and 0.15cm);
    \end{axis}
    \end{tikzpicture}
  \leavevmode \epsfxsize=0.22\textwidth
        \label{5.2}
 }
    \caption{\small{The optimal set ${\lbrace A_m^* +B_m^* \rbrace}^{200}_{m=1}$ with respect to the content object $m$.}}
    \label{5}
\end{figure}
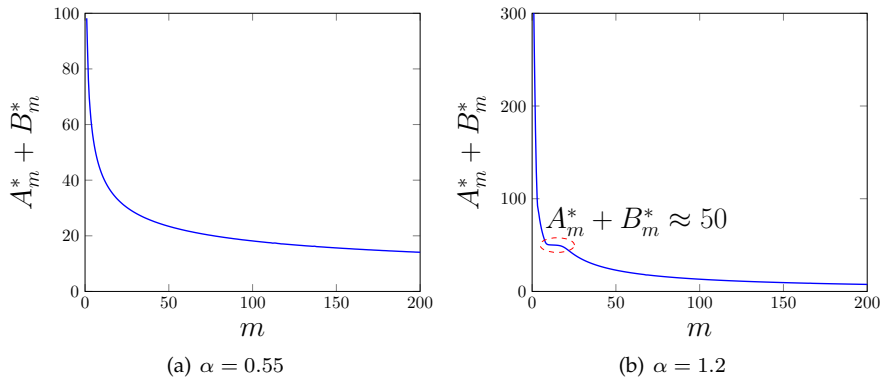

In Fig. \ref{5}, the set $\lbrace A_m^*+B_m^*\rbrace_{m=1}^{200}$
is illustrated according to the content object $m$. It is found
that the simulation results are consistent with our analytical
trend in Theorem \ref{Thm:Theorem2}, which is depicted in Fig. \ref{5a}.
More specifically, for $\alpha =0.55$, it is shown in Fig.
\ref{5.1} that $A_m^*+B_m^*$ monotonically decreases as $m$
increases. Besides, as illustrated in Fig. \ref{5.2}, there exists
a set of content objects $m$ such that $A_m^*+B_m^*$ is
approximately given by $f(n)=50$ for $\alpha =1.2$.

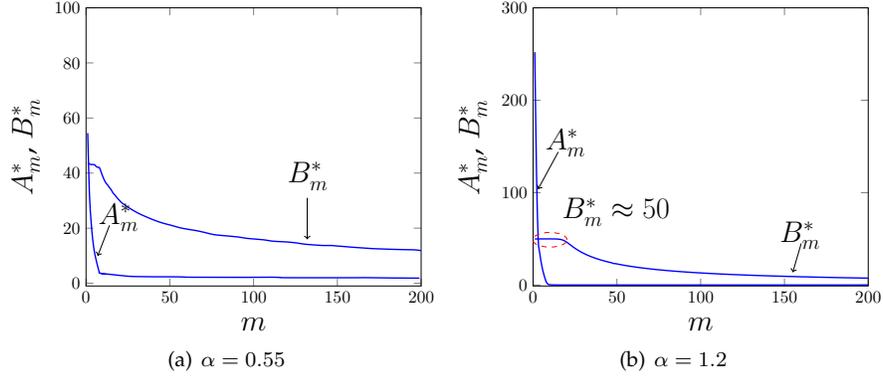
\begin{figure}[t!]
    \centering
     \subfigure[$\alpha =0.55$]{\leavevmode
  \begin{tikzpicture}[scale=0.65]
    \begin{axis}[
        ylabel={\LARGE $A_m^*\textnormal{, } B_m^*$},
        xlabel={\LARGE $m$},xmin=0, xmax=200,ymin=-1, ymax=100]
       \addplot[smooth,blue,thick] plot coordinates {
        (1,54.5)
        (2,32.8)
        (3,22.52)
        (4,15.81)
        (5,11.39)
        (6,8.51)
        (7,6.02)
        (8,3.87)
        (9,3.52)
        (10,3.47)
        (11,3.44)
        (21,2.79)
        (26,2.51)
        (31,2.42)
        (41,2.34)
        (46,2.34)
        (56,2.31)
        (61,2.19)
        (71,2.17)
        (96,2.14)
        (111,2.19)
        (116,2.09)
        (121,2)
        (161,2)
        (166,2.04)
        (181,1.92)
        (199,1.85)
    };
    \addplot[smooth,blue,thick] plot coordinates {
        (1,43.71)
        (2,43.34)
        (3,43.11)
        (5,43.04)
        (6,42.4)
        (7,42.1)
        (8,41.95)
        (9,40.37)
        (10,38.75)
        (11,37.33)
        (12,36.3)
        (13,35.58)
        (14,34.71)
        (15,33.95)
        (16,32.86)
        (17,32.17)
        (21,29.39)
        (26,27.24)
        (31,25.47)
        (36,24.08)
        (41,22.83)
        (46,21.8)
        (51,21.01)
        (56,20.15)
        (61,19.57)
        (66,19.08)
        (71,18.41)
        (76,17.68)
        (81,17.35)
        (91,16.78)
        (96,16.29)
        (106,15.77)
        (111,15.29)
        (116,15.11)
        (121,14.93)
        (126,14.6)
        (131,14.15)
        (141,13.71)
        (146,13.71)
        (151,13.47)
        (171,12.57)
        (191,12.19)
        (196,12.13)
        (200,11.85)
    };
    \draw   (35,25) node[anchor=east] {{\LARGE $A_m^*$}};
    \draw [black, ->          ] (14,22) -- (7,11);
    \draw   (148,40) node[anchor=east] {{\LARGE $B_m^*$}};
    \draw [black, ->          ] (132,32) -- (132,17);
    \end{axis}
    \end{tikzpicture}
  \leavevmode \epsfxsize=0.22\textwidth
       \label{6.1}
}
       \subfigure[$\alpha =1.2$]{\leavevmode
 \begin{tikzpicture}[scale=0.65]
    \begin{axis}[
        ylabel={\LARGE $A_m^*\textnormal{, } B_m^*$},
        xlabel={\LARGE $m$},xmin=0, xmax=200,ymin=-1, ymax=300]
      \addplot[smooth,blue,thick] plot coordinates {
        (1,252.03)
        (2,124.02)
        (3,50.24)
        (4,33.95)
        (5,22.63)
        (6,14.26)
        (7,7.81)
        (8,2.78)
        (9,0.87)
        (10,0.38)
        (11,0.23)
        (12,0.16)
        (13,0.12)
        (14,0.1)
        (15,0.08)
        (16,0.07)
        (17,0.07)
        (18,0.06)
        (19,0.06)
        (20,0.06)
        (21,0.06)
        (22,0.07)
        (23,0.07)
        (24,0.07)
        (25,0.07)
        (26,0.07)
        (27,0.07)
        (28,0.07)
        (29,0.07)
        (30,0.07)
        (31,0.07)
        (32,0.07)
        (33,0.07)
        (34,0.07)
        (35,0.07)
        (36,0.07)
        (37,0.07)
        (38,0.07)
        (39,0.07)
        (40,0.07)
        (41,0.07)
        (42,0.07)
        (43,0.07)
        (44,0.07)
        (45,0.07)
        (46,0.07)
        (47,0.07)
        (48,0.07)
        (49,0.07)
        (50,0.07)
        (51,0.07)
        (52,0.07)
        (53,0.07)
        (54,0.07)
        (55,0.08)
        (56,0.08)
        (57,0.08)
        (58,0.08)
        (59,0.08)
        (60,0.08)
        (61,0.08)
        (62,0.08)
        (63,0.08)
        (64,0.08)
        (65,0.08)
        (66,0.08)
        (67,0.08)
        (68,0.08)
        (69,0.08)
        (70,0.08)
        (71,0.08)
        (72,0.08)
        (73,0.08)
        (74,0.08)
        (75,0.08)
        (76,0.08)
        (77,0.08)
        (78,0.08)
        (79,0.08)
        (80,0.08)
        (81,0.08)
        (82,0.08)
        (83,0.08)
        (84,0.08)
        (85,0.08)
        (86,0.08)
        (87,0.08)
        (88,0.08)
        (89,0.08)
        (90,0.08)
        (91,0.08)
        (92,0.08)
        (93,0.08)
        (94,0.08)
        (95,0.08)
        (96,0.08)
        (97,0.08)
        (98,0.08)
        (99,0.08)
        (100,0.08)
        (101,0.08)
        (102,0.08)
        (103,0.08)
        (104,0.08)
        (105,0.08)
        (106,0.08)
        (107,0.08)
        (108,0.08)
        (109,0.08)
        (110,0.08)
        (111,0.08)
        (112,0.08)
        (113,0.08)
        (114,0.08)
        (115,0.08)
        (116,0.08)
        (117,0.08)
        (118,0.08)
        (119,0.08)
        (120,0.08)
        (121,0.08)
        (122,0.08)
        (123,0.08)
        (124,0.08)
        (125,0.08)
        (126,0.08)
        (127,0.08)
        (128,0.08)
        (129,0.08)
        (130,0.08)
        (131,0.08)
        (132,0.08)
        (133,0.08)
        (134,0.08)
        (135,0.08)
        (136,0.08)
        (137,0.08)
        (138,0.08)
        (139,0.08)
        (140,0.08)
        (141,0.08)
        (142,0.08)
        (143,0.08)
        (144,0.08)
        (145,0.08)
        (146,0.08)
        (147,0.08)
        (148,0.08)
        (149,0.08)
        (150,0.08)
        (151,0.08)
        (152,0.08)
        (153,0.08)
        (154,0.08)
        (155,0.08)
        (156,0.08)
        (157,0.08)
        (158,0.08)
        (159,0.08)
        (160,0.08)
        (161,0.08)
        (162,0.08)
        (163,0.08)
        (164,0.08)
        (165,0.08)
        (166,0.08)
        (167,0.08)
        (168,0.08)
        (169,0.08)
        (170,0.08)
        (171,0.08)
        (172,0.08)
        (173,0.08)
        (174,0.08)
        (175,0.08)
        (176,0.08)
        (177,0.08)
        (178,0.08)
        (179,0.08)
        (180,0.08)
        (181,0.08)
        (182,0.08)
        (183,0.08)
        (184,0.08)
        (185,0.08)
        (186,0.08)
        (187,0.08)
        (188,0.08)
        (189,0.08)
        (190,0.08)
        (191,0.08)
        (192,0.08)
        (193,0.08)
        (194,0.08)
        (195,0.08)
        (196,0.08)
        (197,0.08)
        (198,0.08)
        (199,0.08)
        (200,0.08)
            };
   \addplot[ smooth,blue,thick] plot coordinates {
        (1,49.88)
        (2,49.92)
        (3,49.92)
        (4,49.93)
        (5,49.93)
        (6,49.93)
        (7,49.94)
        (8,49.96)
        (9,49.96)
        (10,49.96)
        (11,49.95)
        (12,49.93)
        (13,49.89)
        (14,49.82)
        (15,49.69)
        (16,49.47)
        (17,49.09)
        (18,48.47)
        (19,47.58)
        (20,46.45)
        (21,45.15)
        (22,43.77)
        (23,42.4)
        (24,41.06)
        (25,39.79)
        (26,38.59)
        (27,37.46)
        (28,36.4)
        (29,35.4)
        (30,34.46)
        (31,33.57)
        (32,32.72)
        (33,31.93)
        (34,31.18)
        (35,30.46)
        (36,29.79)
        (37,29.14)
        (38,28.52)
        (39,27.94)
        (40,27.37)
        (41,26.84)
        (42,26.33)
        (43,25.83)
        (44,25.36)
        (45,24.91)
        (46,24.47)
        (47,24.06)
        (48,23.26)
        (49,23.26)
        (50,22.89)
        (51,22.53)
        (52,22.18)
        (53,21.85)
        (54,21.52)
        (55,21.21)
        (56,20.9)
        (57,20.61)
        (58,20.32)
        (59,20.04)
        (60,19.78)
        (61,19.52)
        (62,19.26)
        (63,19.02)
        (64,18.78)
        (65,18.55)
        (66,18.32)
        (67,18.1)
        (68,17.89)
        (69,17.68)
        (70,17.47)
        (71,17.28)
        (72,17.08)
        (73,16.89)
        (74,16.71)
        (75,16.53)
        (76,16.36)
        (77,16.19)
        (78,16.02)
        (79,15.86)
        (80,15.7)
        (81,15.54)
        (82,15.39)
        (83,15.24)
        (84,15.09)
        (85,14.95)
        (86,14.81)
        (87,14.67)
        (88,14.54)
        (89,14.41)
        (90,14.28)
        (91,14.15)
        (92,14.03)
        (93,13.91)
        (94,13.79)
        (95,13.67)
        (96,13.56)
        (97,13.44)
        (98,13.33)
        (99,13.22)
        (100,13.12)
        (101,13.01)
        (102,12.91)
        (103,12.81)
        (104,12.71)
        (105,12.61)
        (106,12.52)
        (107,12.42)
        (108,12.33)
        (109,12.24)
        (110,12.15)
        (111,12.06)
        (112,11.97)
        (113,11.89)
        (114,11.8)
        (115,11.72)
        (116,11.64)
        (117,11.56)
        (118,11.48)
        (119,11.4)
        (120,11.33)
        (121,11.25)
        (122,11.18)
        (123,11.1)
        (124,11.03)
        (125,10.96)
        (126,10.89)
        (127,10.82)
        (128,10.75)
        (129,10.69)
        (130,10.62)
        (131,10.55)
        (132,10.49)
        (133,10.43)
        (134,10.36)
        (135,10.3)
        (136,10.24)
        (137,10.18)
        (138,10.12)
        (139,10.06)
        (140,10)
        (141,9.95)
        (142,9.89)
        (143,9.83)
        (144,9.78)
        (145,9.72)
        (146,9.67)
        (147,9.62)
        (148,9.56)
        (149,9.51)
        (150,9.46)
        (151,9.41)
        (152,9.36)
        (153,9.31)
        (154,9.26)
        (155,9.21)
        (156,9.17)
        (157,9.12)
        (158,9.07)
        (159,9.03)
        (160,8.89)
        (161,8.94)
        (162,8.89)
        (163,8.85)
        (164,8.8)
        (165,8.76)
        (166,8.72)
        (167,8.68)
        (168,8.63)
        (169,8.59)
        (170,8.55)
        (171,8.51)
        (172,8.47)
        (173,8.43)
        (174,8.39)
        (175,8.35)
        (176,8.32)
        (177,8.28)
        (178,8.24)
        (179,8.2)
        (180,8.16)
        (181,8.13)
        (182,8.09)
        (183,8.06)
        (184,8.02)
        (185,7.99)
        (186,7.95)
        (187,7.92)
        (188,7.88)
        (189,7.85)
        (190,7.82)
        (191,7.78)
        (192,7.75)
        (193,7.72)
        (194,7.69)
        (195,7.65)
        (196,7.62)
        (197,7.59)
        (198,7.56)
        (199,7.53)
        (200,7.5)
            };
   \draw    (35,155) node[anchor=east] {{\LARGE $A_m^*$}};
    \draw [black, ->          ] (15,145) -- (3,105);
    \draw   (175,55) node[anchor=east] {{\LARGE $B_m^*$}};
    \draw [black, ->          ] (160,46) -- (155,15);
    \draw   (85,80) node[anchor=east] {{\LARGE $B_m^*\approx 50$}};
    \draw[dashed,red] (10,50) ellipse (0.35cm and 0.15cm);
    \end{axis}
    \end{tikzpicture}
  \leavevmode \epsfxsize=0.22\textwidth
        \label{6.2}
 }
 \caption{\small{The proposed individual replication sets ${\lbrace A_m^* \rbrace}^{200}_{m=1}$ and ${\lbrace B_m^* \rbrace}^{200}_{m=1}$ with respect to the content object $m$.}}
    \label{6}
\end{figure}

In Fig. \ref{6}, the sets ${\lbrace A_m^* \rbrace}^{200}_{m=1}$
and ${\lbrace B_m^* \rbrace}^{200}_{m=1}$ are illustrated
according to $m$. It is also found that the overall trends in the
figure are similar to our analytical result in Proposition
\ref{cache1}, which is depicted in Fig. \ref{6aa}. Our numerical
results indicate that only a subset of highly popular content
objects are mainly cached and delivered by mobile devices. It is
also observed that $B_m^*$ is close to $f(n)=50$ for small $m$ and
then decreases as $m$ increases.

\section{Baseline Strategy in Mobile Hybrid IoT Networks}

Since our network model has never been studied before, we present our own baseline scheme that employs a rather na\"{\i}ve caching strategy. More specifically, we present a baseline cache
allocation strategy, where the replication sets ${\lbrace A_m
\rbrace}^M_{m=1}$ and ${\lbrace B_m \rbrace}^M_{m=1}$ are found by
solving two separate optimization problems instead of our original
problem in (\ref{equa10}). We first present the problem
formulation and then find the individual replication sets.
Finally, performance comparison is conducted between two cache
allocation strategies.
\subsection{Problem Formulation}
For the baseline approach, we consider two scenarios such that all
content objects are stored {\em only at mobile devices} (i.e.,
$K_{FAP}=0$) or {\em only at static FAPs} (i.e., $K_n=0$). Then, the objective
function (\ref{cond1}) boils down to $\sum_{m=1}^{M}
\frac{p_m}{\sqrt{A_m}}$ or $\sum_{m=1}^{M}
\frac{p_m}{\sqrt{B_m}}$. To be specific, two optimization problems
can be formulated as follows:
\begin{subequations}
\label{P61}
\begin{align}
    \min_{{\lbrace{A_m\rbrace}}^M_{m=1}}
        & \sum_{m=1}^{M} \frac{p_m}{\sqrt{A_m}}  \label{cond61}\\
    \text{subject to}
        & \sum_{m=1}^{M} A_m \leq nK_n,   \label{6a}\\
        & 1\leq A_m \leq n  \text{ for }  m \in \mathcal{M} \, \label{6c}
\end{align}
and
\end{subequations}
\begin{subequations}
\label{P62}
\begin{align}
    \min_{{\lbrace B_m \rbrace}^M_{m=1}}
        & \sum_{m=1}^{M} \frac{p_m}{\sqrt{B_m}}   \label{cond62}\\
    \text{subject to}
        & \sum_{m=1}^{M} B_m \leq f(n)K_{FAP},  \label{6b}\\
        & 1\leq B_m \leq f(n)   \text{ for }  m \in \mathcal{M} \,. \label{6d}
\end{align}
\end{subequations}

In the same manner used in Section \ref{joint}, the discrete variables $A_m$ and $B_m$ are relaxed to real number in $[1,\infty)$ to preserve the convexity of the objective functions (\ref{cond61}) and (\ref{cond62}).

\subsection{Analytical Results}
As in Section \ref{results}, we use the Lagrangian relaxation method for solving two optimization problems in (\ref{P61}) and (\ref{P62}). We establish the following theorem, which characterizes the optimal total number of replicas of content object $m \in \mathcal{M}$.

\begin{thm} \label{cacheB}
Suppose that $\alpha < 3/2$ and the content delivery routing in
Section~\ref{routing} is used for the content-centric mobile hybrid IoT network. Then, the order-optimal solutions to (\ref{P61}) and
(\ref{P62}) lead to
\begin{equation}
A^{*}_m+B^{*}_m=
  \begin{cases}
    \Theta \left(m^{-\frac{2\alpha}{3}}n^{1-\gamma\left(1-\frac{2\alpha}{3}\right)}\right) & \textnormal{for } m \in \mathcal{M}_4, \\
   \Theta\left(n^\delta \right) & \textnormal{for } m\in\mathcal{M}_2\setminus \mathcal{M}_4, \\
   \Theta\left(m^{-\frac{2\alpha}{3}}n^{\beta+\delta-\gamma\left(1-\frac{2\alpha}{3}\right)}\right) & \textnormal{for } m \in\mathcal{M}\setminus \mathcal{M}_2,
  \end{cases}\nonumber
\end{equation}
where $\mathcal{M}_2 = \lbrace 1,...,m_2-1 \rbrace$ and $\mathcal{M}_4 = \lbrace 1,...,m_4-1 \rbrace$. Here, $m_2=\Theta\left(n^{\gamma-(\gamma-\beta)\frac{3}{2\alpha}}\right)$ and $m_4=\Theta\left(n^{\gamma-(\gamma+\delta-1)\frac{3}{2\alpha}}\right)$.
\end{thm}

\begin{IEEEproof}
Since the first problem (\ref{P61}) corresponds to the
optimization problem for the IoT network case with no FAP, the
order-optimal replication set ${\lbrace A_m^* \rbrace}^M_{m=1}$ is given
by
\begin{equation}
\label{equa61}
A^{*}_m=
 \Theta\left(m^{-\frac{2\alpha}{3}}n^{1-\gamma\left(1-\frac{2\alpha}{3}\right)}\right)
\end{equation}
for $\alpha<\frac{3}{2}$ \cite{a1}.

Now, let us turn to solving the second problem (\ref{P62}), which
corresponds to the scenario where content objects are stored only
at FAPs. Using arguments similar to those in the proof of Theorem
\ref{Thm:Theorem2}, it is not difficult to derive that
$B_m^*=\Theta\left(f(n)\right)=\Theta\left(n^\delta\right)$ for $m
\in \mathcal{M}_2 \triangleq \lbrace1,...,m_2-1\rbrace$ and
\begin{equation}
\label{equa65}
B^{*}_m=\Theta\left(m^{-\frac{2\alpha}{3}}n^{\beta+\delta-\gamma\left(1-\frac{2\alpha}{3}\right)}\right)
\end{equation}
for $m \in \mathcal{M}\setminus \mathcal{M}_2$, where $m_2=\Theta\left(n^{\gamma-(\gamma-\beta)\frac{3}{2\alpha}}\right)$.\par
From (\ref{equa61}) and (\ref{equa65}), we next focus on deciding when each replication set $A_m^*$ or $B_m^*$ is dominant for all $m \in \mathcal{M}$. Under the given condition that $\delta+\beta\geq 1$, we have $B^{*}_m=\Omega(A^{*}_m)$ for $m \in \mathcal{M}\setminus \mathcal{M}_2$. Let $\mathcal{M}_4\triangleq\lbrace 1,...,m_4-1 \rbrace$ denote the set of content objects such that $A_m^*+B_m^*=\Theta(A^{*}_m)$ under our baseline strategy, where $m_4-1$ is the largest index of the set $\mathcal{M}_4$. Then, using (\ref{equa61}) and the fact that $A_{m_4-1}^*=\Theta\left(B_{m_4-1}^*\right)=\Theta\left(n^\delta\right)$, we have $n^\delta=  \Theta\left(m_4^{-\frac{2\alpha}{3}}n^{1-\gamma\left(1-\frac{2\alpha}{3}\right)}\right)$, which results in $m_4=\Theta\left(n^{\gamma-(\gamma+\delta-1)\frac{3}{2\alpha}}\right)$. This completes the proof of the theorem.
\end{IEEEproof}
Now, the above baseline strategy is compared with the order-optimal one in Section \ref{joint}. The throughput--delay trade-off of the baseline strategy can be found by applying the result in Theorem \ref{cacheB} to the objective function (\ref{cond1}). As a consequence, we have
\begin{align}\label{equa66}
&\sum_{m=1}^{M} \frac{p_m}{\sqrt{A_m^*+B_m^*}}\nonumber\\&=\!\sum_{m=1}^{m_4-1}\! \frac{p_m}{\sqrt{A_m^*\!+\!B_m^*}}\!+\!\sum_{m=m_4}^{m_2-1}\! \frac{p_m}{\sqrt{A_m^*\!+\!B_m^*}}\!+\!\sum_{m=m_2}^{M}\! \frac{p_m}{\sqrt{A_m^*\!+\!B_m^*}} \nonumber \\&=
\!\Theta\!\!\left(\!\!\frac{n^{\left(\!\gamma\!-\!(\!\gamma\!+\!\delta\!-\!1)\frac{1}{\alpha}\right)(\frac{3}{2}-\alpha)-\frac{1}{2}}}{H_{\alpha}(M)}\!\!\right)\!+\!\Theta\!\!\left(\!\frac{\!n^{\!-\!\frac{\delta}{2}}\!\max\!\lbrace\! H_{\alpha}\!(m_2)\!,\!H_{\alpha}\!(m_4)\!\rbrace\!}{H_{\alpha}(M)}\!\right) \!\!\nonumber \\&+\Theta\!\left(\!\frac{n^{\gamma(\frac{3}{2}\!-\!\alpha)\!-\!\frac{\beta+\delta}{2}}}{H_{\alpha}(M)}\!\right)\!.
\end{align}

Let the first, second, and third terms in the right-hand side of
(\ref{equa66}) be denoted by $F_4$, $F_5$, and $F_3$,
respectively. Then, it is seen that $F_5=\Theta(F_4)$ for
$1<\alpha<\frac{3}{2}$ and $F_5=O(F_3)$ for $\alpha\leq 1$. Thus,
by comparing the relative size of $F_3$ and $F_4$, one can show
that if
$\frac{3(\gamma+\delta-1)}{3(\gamma+\delta-1)-(\gamma-\beta)}\leq
\alpha < \frac{3}{2}$, then $F_4=\Omega\left(F_3\right)$; and
$F_4=o\left(F_3\right)$ otherwise. Next, for comparison with the
order-optimal strategy in Section \ref{joint}, we take into account the
following three cases according to the values of $\alpha$. For
$\alpha<\frac{3(\gamma+\delta-1)}{3(\gamma+\delta-1)-(\gamma-\beta)}$,
the objective function (\ref{cond1}) scales as $F_3$, which
implies that both order-optimal and baseline strategies have the same
throughput--delay trade-off. For
$\frac{3(\gamma+\delta-1)}{3(\gamma+\delta-1)-(\gamma-\beta)}\leq
\alpha< 1+\frac{\gamma-\beta}{2(\gamma+\delta-1)}$ and
$1+\frac{\gamma-\beta}{2(\gamma+\delta-1)}\leq \alpha<
\frac{3}{2}$, (\ref{cond1}) scales as $F_3$ and $F_1$,
respectively, under the order-optimal strategy, while it scales as $F_4$
under the baseline. Since $F_4$ is greater than both $F_3$ and
$F_1$ in these two cases, the order-optimal one provides a better
throughput--delay trade-off for
$\frac{3(\gamma+\delta-1)}{3(\gamma+\delta-1)-(\gamma-\beta)}\leq
\alpha< \frac{3}{2}$. Therefore, the baseline cache allocation
strategy is strictly suboptimal in terms of scaling laws.

\begin{remark}
The performance gain over the baseline approach basically comes
from the fact that in our order-optimal cache allocation strategy,
mobile devices can further cache highly popular content objects,
whose number of replicas is larger than the number of static FAPs. This
finally leads to the performance improvement in terms of
throughput--delay trade-off when
$A_m^*+B_m^*=\omega\left(n^\delta\right)$.
\end{remark}

Note that for $\alpha\geq\frac{3}{2}$ (corresponding to the case
where the use of FAPs is not beneficial), the objective function
(\ref{cond1}) scales as $\frac{1}{\sqrt{n}}$ when the baseline
strategy is used, which also exhibits the best throughput--delay
trade-off in our network from the result of Theorem
\ref{tradeoff}.

\section{Concluding Remarks}
This paper analyzed the order-optimal throughput--delay trade-off in
content-centric mobile hybrid IoT networks, where each of mobile devices
and static FAPs is equipped with a finite cache size. A content
delivery routing was first proposed to characterize a fundamental
throughput--delay trade-off. Then, the order-optimal cache allocation
strategy, which jointly finds the number of replicas cached at
mobile devices and static FAPs using a variable decoupling technique, was
presented to achieve the order-optimal throughput--delay trade-off. Our
analytical results were comprehensively validated by numerical
evaluation.

Suggestions for further research includes characterizing the
optimal throughput--delay trade-off in mobile IoT networks for
the case where the size of each content object is considerably
large and thus the per-hop time required for content delivery may
be greater than the duration of each time slot.

\appendices
\renewcommand\theequation{\Alph{section}.\arabic{equation}}
\setcounter{equation}{0}

\section{Proof of Lemma~\ref{m2}}\label{PF:Lemma2}

Let us first prove that $A_m^*$ is non-increasing with $m \in
\mathcal{M}_1$. From (\ref{AAA}), we have
 \begin{equation}
\label{L31} -\frac{p_m}{2\sqrt{A_m^{*3}}} + \lambda^*+w_m^*=0
\textnormal{ for } m \in \mathcal{M}_1.
\end{equation}
Let $\mathcal{D}_1\subset\mathcal{M}_1$ denote the set of content
objects such that $A_m^*=n$ and $m_0$ denote the smallest index
such that $A_{m_0}^*<n$. Now, consider any content object
$k\in\mathcal{D}_1$. Then, using (\ref{L31}) and the fact that
$w_{m_0}^*=0$, we have
$\lambda^*=\frac{p_{k}}{2\sqrt{A_{k}^{*3}}}-w_{k}^*=\frac{p_{m_0}}{2\sqrt{A_{m_0}^{*3}}}>0$.
Since $A^*_{k}=n$, $A^*_{m_0}<n$, and $w^*_k  \geq 0$, we obtain
$p_{k}>p_{m_0}$, thus resulting in $k<m_0$ due to the feature of a
Zipf popularity in (\ref{eq1}). That is,
$\mathcal{D}_1=\{1,2,...,m_0-1\}$. Furthermore, for $m \in
\mathcal{M}_1\setminus \mathcal{D}_1$, using (\ref{A}) and
(\ref{L31}), we have
 $A_m^* =\frac{ p^{\frac{2}{3}}_m}{(2\lambda^*)^{\frac{2}{3}}}$,
which decreases with $m$. Therefore, $A_m^*$ is non-increasing
with $m\in\mathcal{M}_1$.

Let us turn to proving that $B_m^*$ is non-increasing with $m \in
\mathcal{M}_3$. Let $\mathcal{D}_2\subset\mathcal{M}_3$ denote the
set of content objects such that $B_m^*=f(n)$ and $\tilde{m}_0$
denote the smallest index such that $B_{\tilde{m}_0}^*<f(n)$. By
applying the KKT conditions for (\ref{equa110}) and the same
approach as above, one can show that
$\mathcal{D}_2=\{1,2,\cdots,\tilde{m}_0-1\}$ and
$B_m^*=\frac{p_m^{\frac{2}{3}}}{(2\mu^*)^{\frac{2}{3}}}$ for
$m\in\mathcal{M}_3\setminus \mathcal{D}_2$, which decreases with
$m$. Therefore, $B_m^*$ is non-increasing with
$m\in\mathcal{M}_3$. This completes the proof of the lemma.

\section{Proof of Theorem~\ref{Thm:Theorem2}}\label{PF:Theorem2}

As shown in the proof of Lemma~\ref{m2} (see
Appendix~\ref{PF:Lemma2}), applying the KKT conditions for
(\ref{equa1100}) and (\ref{equa110}), we have
\begin{align}
\label{L321}
 A_m^* =\frac{ p^{\frac{2}{3}}_m}{(2\lambda^*)^{\frac{2}{3}}}~\text{for}~m \in \mathcal{M}_1\setminus \mathcal{D}_1
\end{align}
and
\begin{align}
\label{L36}
   B_m^* =\frac{ p^{\frac{2}{3}}_m}{(2\mu^*)^{\frac{2}{3}}}~\text{for}~m \in \mathcal{M}_3\setminus \mathcal{D}_2,
\end{align}
where $\mathcal{D}_1\subset\mathcal{M}_1$ and
$\mathcal{D}_2\subset\mathcal{M}_3$ are the sets of content
objects such that $A_m^*=n$ and $B_m^*=f(n)$, respectively. From
the proof of Lemma \ref{m2}, recall that $\mathcal{D}_1=\lbrace
1,...,m_0-1\rbrace$, where $m_0$ denotes the smallest index such
that $A_{m}^*<n$. Using (\ref{AA}) and (\ref{BBB}), one can show
that $\sum_{m=1}^M A_m^* = nK_{n}$ and $\sum_{m=1}^M B_m^* =
f(n)K_{FAP}$ due to the fact that $\lambda^*>0$ and $\mu^*>0$,
respectively, which will be used for computing the sum of
$A_m^*+B_m^*$ over some indices in $\mathcal{M}$.

The optimization problem (\ref{equa10}) can be solved according to
the following two cases depending on how the Lagrangian multiplier
$\lambda^*$ in (\ref{equa13}) scales with the another one $\mu^*$
in (\ref{equa13b}). In particular, we consider two cases where
$\lambda^*=\Theta(\mu^*)$ and
$\lambda^*\neq\Theta\left(\mu^*\right)$, each of which corresponds
to the cases where $\alpha \leq
\frac{3(\gamma-\beta)}{2(\delta+\gamma-1)}$ and
$\frac{3(\gamma-\beta)}{2(\delta+\gamma-1)}<\alpha<\frac{3}{2}$.

{\bf Case 1}: We first consider the case where
$\lambda^*=\Theta(\mu^*)$.
Using (\ref{L321}) and (\ref{L36}), one can show that
$A_m^*+B_m^*$ is computed as
\begin{align}
\label{L361}
   A_m^*+B_m^* =\frac{ p^{\frac{2}{3}}_m}{\xi^{\frac{2}{3}}},
\end{align}
where $\xi=\Theta(\lambda^*)=\Theta(\mu^*)$ for $\forall m \in
\mathcal{M}\setminus \mathcal{D}_1$. By adding up $A_m^*+B_m^*$ in
(\ref{L361}) over $\forall m \in
\mathcal{M}\setminus\mathcal{D}_1$ and using the fact that
$\xi^{\frac{2}{3}} =\frac{
\sum_{l=m_0}^{M}p^{\frac{2}{3}}_l}{\sum_{l=m_0}^{M}(A_l^* +
B_l^*)}$, we obtain
\begin{align}
\label{L39} A_m^* + B_m^* &=\frac{
p^{\frac{2}{3}}_m}{\sum_{l=m_0}^{M}p^{\frac{2}{3}}_l}\sum_{l=m_0}^{M}(A_l^*
+
B_l^*)\nonumber\\&=\Theta\left(m^{-\frac{2\alpha}{3}}n^{\beta+\delta-\gamma\left(1-\frac{2\alpha}{3}\right)}\right).
\end{align}
Here, the second equality holds since the cardinality of
$\mathcal{D}_1$ is $O(1)$ due to the cache space $K_n=\Theta(1)$;
$\sum_{l=m_0}^{M}(A_l^* + B_l^*)=\Theta(n^{\delta+\beta})$ from
the assumption that $\delta+\beta\geq 1$ as
$K_{FAP}=\Theta\left(n^\beta\right)$ and
$f(n)=\Theta\left(n^\delta\right)$; from (\ref{eq1}) and
(\ref{equa9}), $\sum_{l=m_0}^{M}
p^{\frac{2}{3}}_l=\sum_{l=m_0}^{M}
\frac{l^{-\frac{2\alpha}{3}}}{H^{\frac{2}{3}}_{\alpha}(M)}=\Theta\left(\frac{M^{1-\frac{2\alpha}{3}}}{H^{\frac{2}{3}}_{\alpha}(M)}\right)$
for $\alpha<\frac{3}{2}$; and $M=\Theta\left(n^\gamma\right)$.\par
Now, we proceed with proving that the set $\mathcal{D}_1$ does not
exist by contradiction. Suppose that there exists $\mathcal{D}_1$
(or equivalently $m_0>1$). When the largest index in the set
$\mathcal{M}_2$ of content objects such that
$A_m^*+B_m^*=\Omega(f(n))$ is denoted as $m_2-1$, it follows that
$A_{m_2-1}^*+B_{m_2-1}^*=\Theta(f(n))$. Using (\ref{L39}), we have
$f(n)=\Theta\left((m_2-1)^{-\frac{2\alpha}{3}}n^{\beta+\delta-\gamma\left(1-\frac{2\alpha}{3}\right)}\right)$,
which then yields
\begin{equation}
\label{L390}
m_2=\Theta\left(n^{\gamma-(\gamma-\beta)\frac{3}{2\alpha}}\right).
\end{equation}
Meanwhile, it is seen that
$\mathcal{M}_1\setminus\mathcal{D}_1=\lbrace
m_0,...,m_2-1\rbrace$. This is because the set $\mathcal{M}_2$
belongs to $\mathcal{M}_1$ from the fact that $B_m^*\leq f(n)$. By
adding up $A_m^*+B_m^*$ in (\ref{L361}) over all $m \in
\mathcal{M}_1\setminus\mathcal{D}_1$, we thus have
\begin{align}
A_{m}^*+B_{m}^*=\Theta\left(A_{m}^*\right)=\frac{
p^{\frac{2}{3}}_{m}}{\sum_{l=m_0}^{m_2-1}p^{\frac{2}{3}}_l}\sum_{l=m_0}^{m_2-1}\Theta\left(A_l^*\right),\nonumber
\end{align}
which results in
\begin{align}
\label{L397} A_{m_0}^*
=\Theta\left(\frac{\sum_{l=m_0}^{m_2-1}A_l^*}{(m_2-1)^{1-\frac{2\alpha}{3}}}\right)
\end{align}
when $m=m_0$ due to the fact that from (\ref{eq1}) and
(\ref{equa9}), $\sum_{l=m_0}^{m_2-1}
p^{\frac{2}{3}}_l=\sum_{l=m_0}^{m_2-1}
\frac{l^{-\frac{2\alpha}{3}}}{H^{\frac{2}{3}}_{\alpha}(M)}=\Theta\left(\frac{(m_2-1)^{1-\frac{2\alpha}{3}}}{H^{\frac{2}{3}}_{\alpha}(M)}\right)$
for $\alpha<\frac{3}{2}$. Since
$(m_2-1)^{1-\frac{2\alpha}{3}}=\omega(1)$ from (\ref{L390}) and
$\sum_{l=m_0}^{m_2-1}A_l^*=O(n)$, it follows that $A_{m_0}^*=o(n)$
from (\ref{L397}), which contradicts another relation
$A_{m_0}^*=\Theta(n)$. Hence, we conclude that $\mathcal{D}_1$
does not exist (i.e., $m_0=1$).\par In the following, we show that
the condition $\lambda^*=\Theta\left(\mu^*\right)$ is identical to
$\alpha \leq \frac{3(\gamma-\beta)}{2(\delta+\gamma-1)}$. Using
(\ref{L39}) and (\ref{L390}) as well as $m_0=1$, we have
$\sum_{m=1}^{m_2-1}(A_m^* + B_m^*) =\frac{
\sum_{m=1}^{m_2-1}p^{\frac{2}{3}}_m}{\sum_{l=1}^{M}p^{\frac{2}{3}}_l}\sum_{l=1}^{M}(A_l^*
+
B_l^*)=\Theta\left(n^{\delta+\beta-(\gamma-\beta)\left(\frac{3}{2\alpha}-1\right)}\right)$.
Using the fact that $\sum_{m=1}^{m_2-1}(A_m^* +
B_m^*)=\sum_{m=1}^{m_2-1}\Theta\left(A_m^*\right)=O(n)$, we obtain
the following inequality:
$\delta+\beta-(\gamma-\beta)\left(\frac{3}{2\alpha}-1\right)\leq
1$, which is equivalent to
$\alpha \leq \frac{3(\gamma-\beta)}{2(\delta+\gamma-1)}$.
Therefore, if $\alpha \leq
\frac{3(\gamma-\beta)}{2(\delta+\gamma-1)}$, then
\begin{equation}
\label{equa12}
A^{*}_m+B^{*}_m=
 \Theta\left(m^{-\frac{2\alpha}{3}}n^{\beta+\delta-\gamma\left(1-\frac{2\alpha}{3}\right)}\right) \text{ for }  m \in \mathcal{M}.\nonumber
\end{equation}

{\bf Case 2}: We turn our attention to the case where
$\lambda^*\neq\Theta\left(\mu^*\right)$. From (\ref{L321}) and
(\ref{L36}), it is found that the two sets $\mathcal{M}_1\setminus
\mathcal{D}_1$ and $\mathcal{M}_3\setminus \mathcal{D}_2$ do not
share any common content object. Thus, it follows that
$\mathcal{M}_1\subset\mathcal{M}_2$. Based on this observation, we
have
\begin{align}
\label{T1} A_m^*+B_m^*=\Theta(B_m^*)=\Theta\left(n^\delta\right)
\end{align}
for $m \in \mathcal{M}_2\setminus\mathcal{M}_1$. For the sake of
analytical convenience, by defining $m_1-1$ and $m_2-1$ as the
largest indices in the sets $\mathcal{M}_1$ and $\mathcal{M}_2$,
respectively, we compute $A_m^*+B_m^*$ for $m\in\mathcal{M}_1$ and
$m\in\mathcal{M}\setminus\mathcal{M}_2$.

First, we focus on finding both $A_m^*+B_m^*$ and $m_2$ for $m
\in\mathcal{M}\setminus\mathcal{M}_2$, where
$\mathcal{M}\setminus\mathcal{M}_2=\lbrace m_2,...,M\rbrace$. By
adding up $B_m^*$ in (\ref{L36}) over all $m \in
\mathcal{M}\setminus \mathcal{M}_2$, we have
\begin{subequations}
\begin{align}
B_m^* &=\frac{
p^{\frac{2}{3}}_m}{\sum_{l=m_2}^{M}p^{\frac{2}{3}}_l}\sum_{l=m_2}^{M}
B_l^*\label{L393}\\&=O\left(m^{-\frac{2\alpha}{3}}n^{\delta+\beta-\gamma\left(1-\frac{2\alpha}{3}\right)}\right),\label{L3931}
\end{align}
\end{subequations}
where the second equality holds since
$\sum_{l=m_2}^{M}B_l^*=O\left(n^{\delta+\beta}\right)$; from
(\ref{eq1}) and (\ref{equa9}), $\sum_{l=m_2}^{M}
p^{\frac{2}{3}}_l=\sum_{l=m_2}^{M}
\frac{l^{-\frac{2\alpha}{3}}}{H^{\frac{2}{3}}_{\alpha}(M)}=\Theta\left(\frac{M^{1-\frac{2\alpha}{3}}}{H^{\frac{2}{3}}_{\alpha}(M)}\right)$
for $\alpha<\frac{3}{2}$; and $M=\Theta\left(n^\gamma\right)$.
From (\ref{L3931}) and $B_{m_2}^*=\Theta(f(n))$, we have
$f(n)=O\left(m_2^{-\frac{2\alpha}{3}}n^{\delta+\beta-\gamma\left(1-\frac{2\alpha}{3}\right)}\right),$
which leads to $
m_2=O\left(n^{\gamma~-~(\gamma-\beta)~\frac{3}{2\alpha}}\right)$.
Since $(\gamma~-~\beta)~(1-\frac{3}{2\alpha})<0$ under our network
model, it follows that
$\gamma-(\gamma-\beta)\frac{3}{2\alpha}<\beta$, thus resulting in
$m_2=o(K_{FAP})$. Because
$\sum_{l=m_2}^{M}B_l^*=f(n)K_{FAP}-O(m_2f(n))=\Theta(n^{\delta+\beta})$,
(\ref{L393}) can be rewritten as
\begin{align}
\label{T2} A_m^*+B_m^*=\Theta\left(
B_m^*\right)=\Theta\left(m^{-\frac{2\alpha}{3}}n^{\delta+\beta-\gamma\left(1-\frac{2\alpha}{3}\right)}\right)
\end{align}
for $m\in \mathcal{M}\setminus\mathcal{M}_2$. Moreover, using the
fact that $A_{m_2}^*+B_{m_2}^*
=\Theta\left(m_2^{-\frac{2\alpha}{3}}n^{\delta+\beta-\gamma\left(1-\frac{2\alpha}{3}\right)}\right)$
from (\ref{T2}) and $A_{m_2}^*+B_{m_2}^*=\Theta(f(n))$, we obtain
\begin{align}
\label{L3955}
m_2=\Theta\left(n^{\gamma-(\gamma-\beta)\frac{3}{2\alpha}}\right).
\end{align}
Next, we turn to finding both $A_m^*+B_m^*$ and $m_1$ for $m
\in\mathcal{M}_1$. By adding up $A_m^*$ in (\ref{L321}) over
$\forall m \in \mathcal{M}_1\setminus\mathcal{D}_1$ ($=\lbrace
m_0,...,m_1-1 \rbrace$), we have
\begin{equation}
\label{L3972} A_{m}^* =\frac{
p^{\frac{2}{3}}_{m}}{\sum_{l=m_0}^{m_1-1}p^{\frac{2}{3}}_l}\sum_{l=m_0}^{m_1-1}A_l^*.
\end{equation}
Similarly as in Case 1, we are capable of proving that
$\mathcal{D}_1$ does not exist (i.e., $m_0=1$). Because
$f(n)=\Theta\left(\frac{m_1^{-\frac{2\alpha}{3}}\sum_{l=1}^{m_1-1}A_l^*}{(m_1-1)^{1-\frac{2\alpha}{3}}}\right)$
due to $\sum_{l=1}^{m_1-1}
p^{\frac{2}{3}}_l=\Theta\left(\frac{(m_1-1)^{1-\frac{2\alpha}{3}}}{H^{\frac{2}{3}}_{\alpha}(M)}\right)$
and $A_{m_1-1}^*=\Theta\left(f(n)\right)$, it is not difficult to
show that
\begin{align}
\label{L39551}
m_1=\Theta\left(\frac{\sum_{l=1}^{m_1-1}A_l^*}{f(n)}\right).
\end{align}
Now, we need to specify the term $\sum_{l=1}^{m_1-1}A_l^*$ to find
$m_1$. For $\alpha<\frac{3}{2}$, by setting $m_0=1$, the objective
function (\ref{cond1}) can be expressed as
\begin{align}
\label{equa345}& \sum_{m\!=\!1}^{M}\!
\frac{p_m}{\sqrt{A_m^*\!+\!B_m^*}}\nonumber\\&=\sum_{m=1}^{m_1\!-\!1}\!
\frac{p_m}{\sqrt{A_m^*\!+\!B_m^*}}+\sum_{m\!=\!m_1}^{m_2\!-\!1}
\!\frac{p_m}{\sqrt{A_m^*\!+\!B_m^*}}+\sum_{m\!=\!m_2}^{M}\!
\frac{p_m}{\sqrt{A_m^*\!+\!B_m^*}} \nonumber \\&=
\Theta\!\!\left(\!\!\frac{\left(\!\sum_{m=1}^{m_1\!-\!1}
\!p_{m}^{\frac{2}{3}}\!\!\right)^{\!\frac{3}{2}}}{\sqrt{\sum_{m=1}^{m_1\!-\!1}
\!A_m^*}}\!\right)\!+\!\Theta\!\!\left(\!\!\frac{\sum_{m=m_1}^{m_2\!-\!1}
\!p_{m}}{\sqrt{f(n)}}\!\!\right)\!+\!\Theta\!\!\left(\!\!\frac{\!\left(\!\sum_{m=m_2}^{M}\!
p_{m}^{\frac{2}{3}}\!\right)^{\!\frac{3}{2}}}{\sqrt{\sum_{m=m_2}^{M}
B_m^*}}\!\!\right)\nonumber \\&=
\!\Theta\!\left(\!\frac{m_1^{1-\alpha}}{H_{\alpha}\!(\!M\!)\sqrt{
f(n)}}\!\right)\!+\!\Theta\!\left(\!\frac{\max \lbrace
H_{\alpha}\!(\!m_1\!),H_{\alpha}\!(\!m_2-1\!)\rbrace}{H_{\alpha}(M)\sqrt{f(n)}}\!\right)\!\nonumber\\&+\!\Theta\!\left(\!\frac{M^{3/2-\alpha}}{H_{\alpha}(M)\!\sqrt{\!f(n)\!K_{FAP}}}\!\right)\!,
\end{align}
where the second equality follows from (\ref{T1}) for $m \in
\mathcal{M}_2\setminus\mathcal{M}_1$, (\ref{L393}) for $m \in
\mathcal{M}\setminus\mathcal{M}_2$, and (\ref{L3972}) for $m \in
\mathcal{M}_1$; the third equality holds since $m_2=o(M)$ from
(\ref{L3955}) as well as $\sum_{l=1}^{m_1-1}
p^{\frac{2}{3}}_l=\Theta\left(\frac{(m_1-1)^{1-\frac{2\alpha}{3}}}{H^{\frac{2}{3}}_{\alpha}(M)}\right)$
and $\sum_{l=m_2}^{M}
p^{\frac{2}{3}}_l=\Theta\left(\frac{M^{1-\frac{2\alpha}{3}}}{H^{\frac{2}{3}}_{\alpha}(M)}\right)$
from (\ref{eq1}) and (\ref{equa9}). One can show that the second
term in (\ref{equa345}) scales slower than the other two terms
because for $\alpha\leq 1$, (\ref{cond1}) is given by the third
term in (\ref{equa345}) and for $1<\alpha<\frac{3}{2}$,
(\ref{cond1}) is dominated by either the first or the third term
in (\ref{equa345}). Since for $1<\alpha<\frac{3}{2}$, the first
term (including $m_1$) can be minimized when $m_1$ is maximized,
it follows that $\sum_{l=1}^{m_1-1}A_l^*=\Theta(n)$ from
(\ref{equa1}) (the total caching constraint) and (\ref{L39551}).
Accordingly, we have
\begin{equation}
\label{T3} m_1=\Theta(n^{1-\delta}). \nonumber
\end{equation}
Moreover by setting  $m_0=1$, $\sum_{l=1}^{m_1-1}
p^{\frac{2}{3}}_l=\Theta\left(\frac{m_1^{1-\frac{2\alpha}{3}}}{H^{\frac{2}{3}}_{\alpha}(M)}\right)$,
and $\sum_{l=1}^{m_1-1}A_l^*=\Theta(n)$, (\ref{L3972}) can be
rewritten as
\begin{align}
\label{T4} A_{m}^*+B_m^*=\Theta\left(
A_m^*\right)=\Theta\left(m^{-\frac{2\alpha}{3}}n^{\delta+(1-\delta)\frac{2\alpha}{3}}\right)~\text{for}~m
\in \mathcal{M}_1. \nonumber
\end{align}

As the last case, for $m\in\mathcal{M}_2\setminus \mathcal{M}_1$,
$A_m^*+B_m^*$ is obviously given by $\Theta(n^{\delta})$ from the
definitions of $\mathcal{M}_1$ and $\mathcal{M}_2$.

Finally, from the last paragraph of Case 1, it is obvious that
$\lambda^*\neq\Theta\left(\mu^*\right)$ when
$\frac{3(\gamma-\beta)}{2(\delta+\gamma-1)}<\alpha<\frac{3}{2}$.
This completes the proof of the theorem.



\ifCLASSOPTIONcaptionsoff
  \newpage
\fi


\begin{thebibliography}{99}
\bibitem{c7}T.-A. Do, S.-W. Jeon, and W.-Y. Shin, ``Caching in mobile HetNets: A throughput-delay trade-off perspective," in \emph{Proc. IEEE Int. Symp. Inf. Theory (ISIT)}, Barcelona, Spain, Jul. 2016, pp. 1247-1251.
\bibitem{singlehop}M. Ji, G. Caire, and A. F. Molisch, ``Wireless device-to-device caching networks: Basic principles and system performance," \emph{IEEE J. Sel. Areas Commun.}, vol. 34, no. 1, pp. 176--189, Jan. 2016.
\bibitem{surv} M. Agiwal, A. Roy, and N. Saxena, ``Next generation 5G wireless networks: A comprehensive survey," \emph{IEEE Commun. Surveys Tuts.,} vol. 18, no. 3, pp. 1617--1655, Third Quart., 2016.
\bibitem{acc1}X. Li, X. Wang, K. Li, and V. C. M. Leung, ``CaaS: Caching as a service for 5G networks," \emph{IEEE Access}, vol. 5, pp.  5982--5993, Mar. 2017.
\bibitem{acc2}S. Wang, X. Zhang, Y. Zhang, L. Wang, J. Yang, and W. Wang, ``A survey on mobile edge networks: Convergence of computing, caching and communications," \emph{IEEE Access}, vol. 5, pp. 6757--6779, Mar. 2017.
\bibitem{c8} P. Gupta and P. R. Kumar, ``The capacity of wireless networks," \emph{IEEE Trans. Inf. Theory}, vol. 46, no. 2, pp. 388--404, Mar. 2000.
\bibitem{Franceschetti} M. Franceschetti, O. Dousse, D. N. C. Tse, and P. Thiran, ``Closing the gap in the capacity of wireless networks via percolation theory," \emph{IEEE Trans. Inf. Theory}, vol. 53, no. 3, pp. 1009--1018, Mar. 2007.
\bibitem{GuptaKumar2003} P. Gupta and P. R. Kumar, ``Towards an information theory of large networks: An achievable rate region," \emph{IEEE Trans. Inf. Theory}, vol. 49, no. 8, pp. 1877--1894, Aug. 2003.
\bibitem{Xue} F. Xue, L.-L. Xie, and P. R. Kumar, ``The transport capacity of wireless networks over fading channels," \emph{IEEE Trans. Inf. Theory}, vol. 51, no. 3, pp. 834--847, Mar. 2005.
\bibitem{Shin} W.-Y. Shin, S.-Y. Chung, and Y. H. Lee, ``Parallel opportunistic routing in wireless networks," \emph{IEEE Trans. Inf. Theory}, vol. 59, no. 10, pp. 6290--6300, Oct. 2013.
\bibitem{c11} A. \"{O}zg\"{u}r, O. L\'{e}v\^{e}que, and D. N. C. Tse, ``Hierarchical cooperation achieves optimal capacity scaling in ad hoc networks," \emph{IEEE Trans. Inf. Theory}, vol. 53, no. 10, pp. 3549--3572, Oct. 2007.
\bibitem{c12} M. Grossglauser and D. N. C. Tse, ``Mobility increases the capacity of ad hoc wireless networks," \emph{IEEE/ACM Trans. Netw.}, vol. 10, no. 4, pp. 477--486, Aug. 2002.
\bibitem{c1}A. El Gamal, J. Mammen, B. Prabhakar, and D. Shah, ``Optimal throughput--delay scaling in wireless networks--Part I: The fluid model," \emph{IEEE Trans. Inf. Theory}, vol. 52, no. 6, pp. 2568--2592, Jun. 2006.
\bibitem{Zhang} G. Zhang, Y. Xu, X. Wang, and M. Guizani, ``Capacity of hybrid wireless networks with directional antennas and delay constraint," \emph{IEEE Trans. Commun.}, vol. 58, no. 7, pp. 2097--2106, July 2010.
\bibitem{Li} P. Li, C. Zhang, and Y. Fang, ``The capacity of wireless ad hoc networks using directional antennas,"  \emph{IEEE Trans. Mobile Comput.}, vol. 10, no. 10, pp. 1374--1387, Oct. 2011.
\bibitem{Yoon} J. Yoon, W.-Y. Shin, and S.-W. Jeon, ``Elastic routing in ad hoc networks with directional antennas," {\em IEEE Trans. Mobile Comput.}, vol. 16, no. 12, pp. 3334--3346, Dec. 2017.
\bibitem{Liu} B. Liu, Z. Liu, and D. Towsley, ``On the capacity of hybrid wireless networks," in \emph{Proc. IEEE INFOCOM}, San Francisco, CA, Mar./Apr. 2003, pp. 1543--1552.
\bibitem{Shin11} W.-Y. Shin, S.-W. Jeon, N. Devroye, M. H. Vu, S.-Y. Chung, Y. H. Lee, and V. Tarokh,  ``Improved capacity scaling in wireless networks with infrastructure," \emph{IEEE Trans. Inf. Theory}, vol. 57, no. 8, pp. 5088--5102, Aug. 2011.
\bibitem{c2}S. Gitzenis, G. S. Paschos, and L. Tassiulas, ``Asymptotic laws for joint content replication and delivery in wireless networks," \emph{IEEE Trans. Inf. Theory}, vol. 59, no. 5, pp. 2760--2776, May. 2013.
\bibitem{Ji_IT} M. Ji, G. Caire, and A. F. Molisch, `The throughput-outage tradeoff of wireless one-hop caching
networks," \emph{IEEE Trans. Inf. Theory}, vol. 61, no. 12, pp.
6833--6859, Dec. 2015.
\bibitem{c5} S.-W. Jeon, S.-N. Hong, M. Ji, G. Caire, and A. F. Molisch, ``Wireless multihop device-to-device caching networks," \emph{IEEE Trans. Inf. Theory}, vol. 63, no. 3, pp. 1662--1676, Mar. 2017.
\bibitem{c3}G. Alfano, M. Garetto, and E. Leonardi, ``Content-centric wireless networks with limited buffers: When mobility hurts," \emph{IEEE/ACM Trans. Netw.}, vol. 24, no. 1, pp. 299--311, Feb. 2016.
\bibitem{Adeel}A. Malik, S. H. Lim, and W.-Y. Shin, ``On the effects of subpacketization in content-centric mobile networks," \emph{IEEE J. Sel. Areas Commun.}, to appear.
\bibitem{XLiu}X. Liu, K. Zheng, J. Zhao, X.-Y. Liu, X. Wang, and X. Di, ``Information-centric networks with correlated mobility," \emph{IEEE Trans. Veh. Technol.}, vol. 66, no. 5, pp. 4256--4270, May 2017.
\bibitem{acc3}G. Zhang, J. Liu, J. Ren, L. Wang, and J. Zhang, ``Capacity of content-centric hybrid wireless networks," \emph{IEEE Access}, vol. 5, pp. 1449--1459, Feb. 2017.
\bibitem{a1}M. Mahdian and E. Yeh, ``Throughput and delay scaling of content-centric ad hoc and heterogeneous wireless networks," \emph{IEEE/ACM Trans. Netw.}, vol. 25, no. 5, pp. 3030--3043, Aug. 2017.
\bibitem{Nie}M. A. Maddah-Ali and U. Niesen, ``Fundamental limits of caching," \emph{IEEE Trans. Inf. Theory}, vol. 60, no. 5, pp. 2856--2867, May. 2014.
\bibitem{Nie1}M. A. Maddah-Ali and U. Niesen, ``Decentralized coded caching attains order-optimal memory-rate tradeoff," \emph{IEEE/ACM Trans. Netw.}, vol. 23, no. 4, pp. 1029--1040, Aug. 2014.
\bibitem{Nie3}M. A. Maddah-Ali and U. Niesen, ``Coding for caching: Fundamental limits and practical challenges," \emph{IEEE Commun. Mag.}, vol. 54, no. 8, pp. 23--29, Aug. 2016.
\bibitem{c10}D. E. Knuth, ``Big Omicron and big Omega and big Theta," \emph{ACM SIGACT News}, vol. 8, no. 2, pp. 18--24, Apr.–-Jun. 1976.
\bibitem{RW2}L. Zhou, R. Q. Hu, Y. Qian, and H.-H Chen, ``Energy-spectrum efficiency tradeoff for video streaming over mobile ad hoc networks," \emph{IEEE J. Sel. Areas Commun.}, vol. 31, no. 5, pp. 981--991, May. 2013.
\bibitem{c6}C. Fricker, P. Robert, J. Roberts, and N. Sbihi, ``Impact of traffic mix on caching performance in a content-centric network," in \emph{Proc. IEEE INFOCOM Workshop on Emerging Choices in Named-Oriented Netw. (NoMEN)}, Orlando, FL, Mar. 2012, pp. 310--315.
\bibitem{Zipf}S. Lim, W.-C. Lee, G. Cao, and C. R. Das, ``A novel caching scheme for Internet based mobile ad hoc networks," in
\emph{Proc. IEEE Computer Commun. and Netw. (ICCCN)}, Dallas, TX, USA, Oct. 2003, pp. 38--43.
\bibitem{Zipf1}M. X. Goemans, L. Li, V. S. Mirrokni, and M. Tholtan, ``Market sharing games applied to content distribution in ad hoc networks," \emph{IEEE J. Sel. Areas Commun.}, vol. 24, no. 5, pp. 1020--1033, May. 2006.
\end{thebibliography}
\end{document}